\documentclass[11pt, preprint]{aastex}
\usepackage{subfigure}
\let\tablehead\undefined
\let\tabletail\undefined
\usepackage{supertabular}
\def\etal{~et al.}

\title{\bf Enhanced Abundances in Spiral Galaxies of the Pegasus I Cluster}
\author{ Paul Robertson, Gregory A. Shields, and Guillermo A. Blanc}
\affil{Department of Astronomy, The University of Texas, Austin, TX, 78712; paul@astro.as.utexas.edu, shields@astro.as.utexas.edu, gblancm@astro.as.utexas.edu}
\begin{abstract}
We study the influence of cluster environment on the chemical evolution of spiral galaxies in the Pegasus I cluster.   We determine the gas-phase heavy element abundances  of six galaxies in Pegasus derived from H~II region spectra  obtained from  integral-field spectroscopy.  These abundances are analyzed in the context of Virgo, whose spirals are known to show increasing interstellar metallicity as a function of H I deficiency.  The galaxies in the Pegasus cluster, despite its lower density and velocity dispersion, also display gas loss due to ISM-ICM interaction, albeit to a lesser degree.  Based on the abundances of 3 H I deficient spirals and 2 H I normal spirals, we observe a heavy element abundance offset of $+0.13 \pm 0.07$ dex for the H I deficient galaxies.  This abundance differential is consistent with the differential observed in Virgo for galaxies with a similar H~I deficiency, and we observe a correlation between log(O/H) and the H I deficiency parameter DEF for the two clusters analyzed together.  Our results suggest that similar environmental mechanisms are driving the heavy element enhancement in both clusters.
\end{abstract}


\begin{document}


\section{\bf Introduction}
The effect of environment on galaxy evolution has long been a subject of intense research and debate \citep[see review articles by][for a comprehensive discussion]{haynes84,boselli06}.  Known environmental effects include tidal encounters and mergers, altered morphologies, and stripping of gas from disks.  A natural question is whether the cluster environment has significant effects of the chemical evolution of galaxies.  This topic has received increasing attention in the last few years.  \citet{skillman96} explore the effect of environment on chemical evolution for spirals in the Virgo cluster.  Examining H II region spectra from 9 Virgo spirals, they find the three most H I deficient objects to have O/H abundances 0.3-0.5 dex higher than their gas-normal counterparts.  They suggest that the abundance differential results in part from a lack of infall of metal-poor gas into the spirals in the cluster core.  \citet{dors06} fit photoionization models to the Virgo data, confirming the abundance excess for O/H and N/O.  Other studies, involving large-scale spectroscopic surveys \citep{cooper08,ellison09,zhang09}, see a qualitatively similar galactic metallicity dependence on local galaxy density or gas fraction.  \citet{zhang09} analyze a sample of 800 galaxies in the HyperLeda catalog, concluding gas-poor galaxies display higher heavy-element content for a given stellar mass.  \citet{petropoulo11} examine abundances of dwarf and spiral galaxies in the Hercules cluster, finding higher abundances in dwarf galaxies located in relatively dense environments.  

While these results suggest a significant impact of cluster environment on chemical evolution, the number of detailed studies for individual clusters is small. This paper presents, for the Pegasus I cluster, a study analogous to the \citet{skillman96} analysis of the Virgo cluster.  Pegasus I (hereafter referred to as Pegasus) is a low-density, low velocity dispersion cluster of redshift $\sim$ 3900 km/s in the foreground of the Pisces-Perseus supercluster.  It displays weak X-ray emission, primarily concentrated around the two central ellipticals, without additional substructure \citep[][Figure 1]{canizares86}.  Full details of the cluster's location and member galaxies can be found in \citet{levy07}, including a comparison to Virgo in their Table 4.  A comprehensive comparison of the properties of the Pegasus galaxies to other nearby clusters is presented in \citet{solanes01}.

Because the density of Pegasus is so low, the classical ram pressure stripping effect should not cause significant gas loss in the disks of member spirals.  Furthermore, \citet{vigroux89} conclude that the cluster is in the early stages of gravitational collapse, which suggests that any environment-driven evolution should be in an early phase as well.  Nevertheless, \citet{levy07} demonstrate that the Pegasus spirals are in fact experiencing H I loss as they fall into the cluster, suggesting mechanisms other than the classic ram pressure effect \citep{gunn72} may be at work.  In a follow-up study, \citet{rose10} find that star formation in Pegasus galaxies is suppressed with higher H I deficiency.   Given that the cluster environment has already caused noticeable changes in the gas content and star formation of these galaxies, one might expect changes in nebular abundance as well.  Using the VIRUS-P integral-field spectrograph \citep{hill08} on the 2.7m Harlan J Smith Telescope at McDonald Observatory, we have obtained spectra for six Pegasus spirals.  From the H II region spectra, we calculate radial O/H profiles, and examine the extent to which a metallicity offset can be seen between the gas-poor and gas-normal spirals.  We compare our results to those of \citet{skillman96}, noting the difference in environments of Virgo and Pegasus.  We discuss the possible causes of the offset for Pegasus, and the potential implications for environment-driven chemical evolution in Virgo and other clusters.

\section{\bf Sample and Observations}
\subsection{Target Galaxy Selection}
Our targets were selected from among the Pegasus spirals analyzed for H I content in \citet{levy07}.  The sample was chosen so as to cover a wide range of values for DEF, the overall galactic H I deficiency value.  As described in \citet{levy07}, DEF measures an offset between a galaxy's H I content (measured from 21 cm radio emission) and an expectation value based on field galaxies of similar luminosity and morphological type, with increasingly positive values indicating higher gas deficiency.  A value of DEF $\sim$ 0.3 indicates H I is deficient by a factor of two, and is considered the threshold at which a galaxy is considered definitively H I deficient.  

The nature of our observations placed further constraints on our target selection.  In order to obtain adequate radial coverage of our sample, we selected spirals with a sufficient number of bright H II regions, as determined from the H $\alpha$ images obtained in \citet{rose10}.  This requirement limited somewhat the highest DEF values we could explore, since, as discussed in \citet{rose10}, the prominence of H II regions is severely truncated at large DEF.  Additionally, we observed the most face-on objects to minimize inclination and reddening effects.  We note that in the optical images of our target galaxies, we see no nearby companions, tidal tails, or other morphological peculiarities.  Furthermore, the galaxies are not listed as having morphological abnormalities in the literature, and do not have close neighbors in the cluster map of \citet[][Figure 8]{levy07}.  We therefore conclude that our targets are not members of interacting pairs.  However, NGC 7643 is approximately 200 kpc away from UGC 12562, the possible effects of which will be addressed in the discussion.

Table \ref{targettab} lists the names, coordinates, inclination-corrected circular velocities, absolute B magnitudes, and effective/isophotal radii of the six spirals observed in this paper \citep{paturel03,rose10}.  Of these, IC 5309, NGC 7518, and NGC 7643 serve as our hydrogen-deficient sample.  The other three objects--NGC 7591, NGC 7529, and IC 1474--have either normal or high H I content, and serve as the control sample.  For all galaxy data, we assume a cosmology with H$_0 = 73.0$ km s$^{-1}$ Mpc$^{-1}$, $\Omega_m = 0.27$, $\Omega_\Lambda = 0.73$.

\subsection{Observations}
\begin{figure}
          \subfigure[IC 5309]{\includegraphics[scale=0.2]{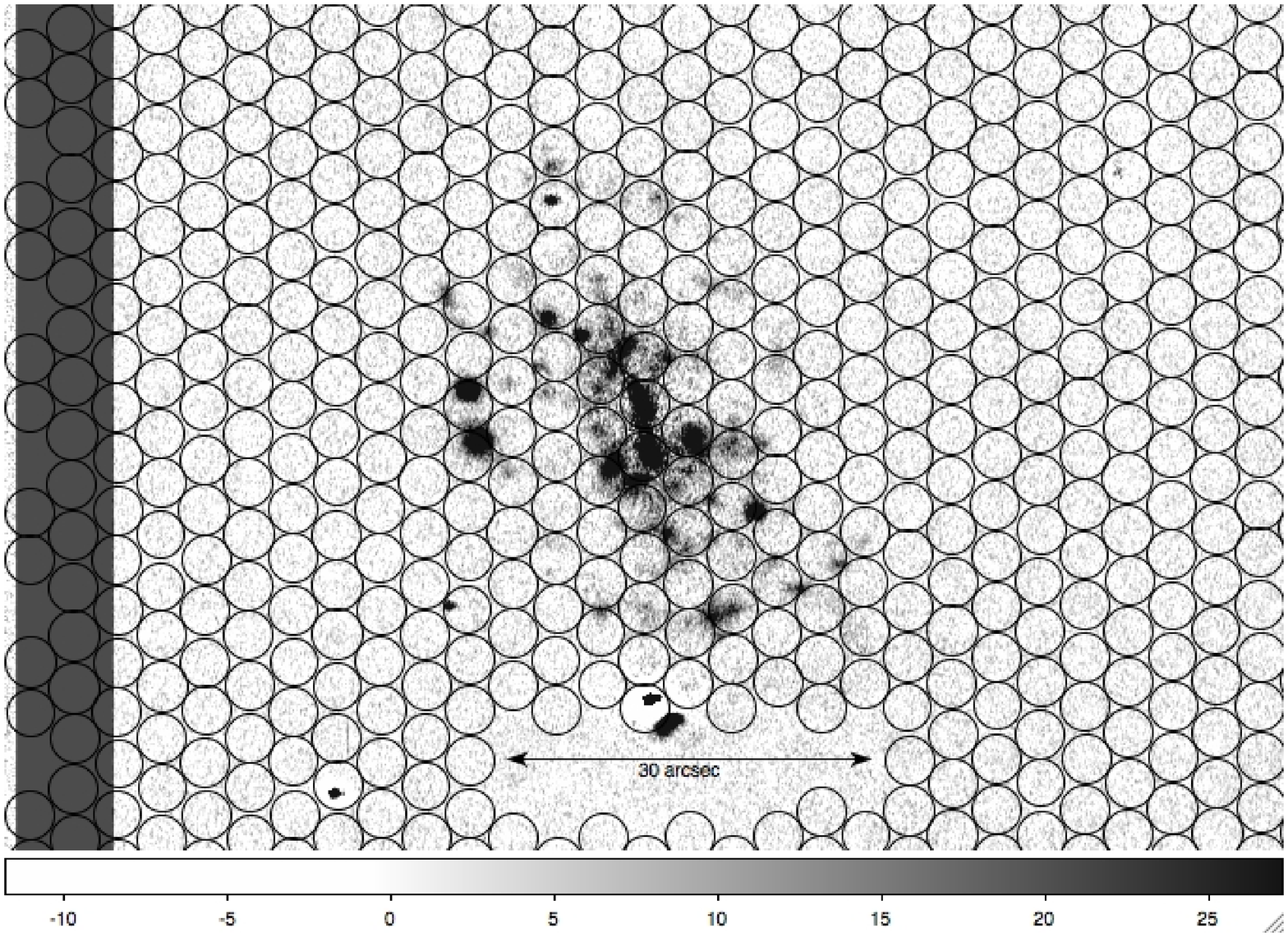}}
          \subfigure[NGC 7643]{\includegraphics[scale=0.2]{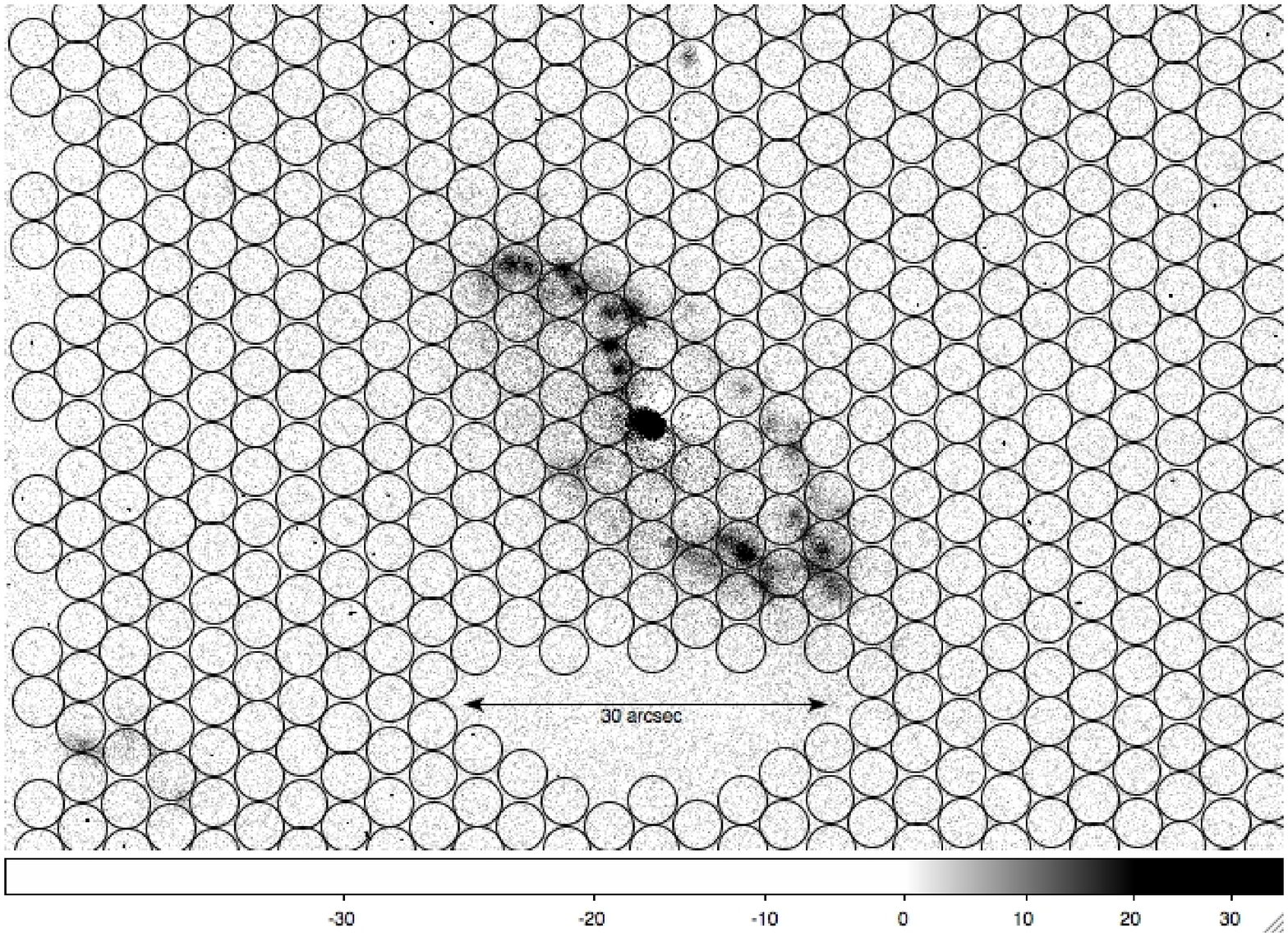}}
          \subfigure[NGC 7518]{\includegraphics[scale=0.2]{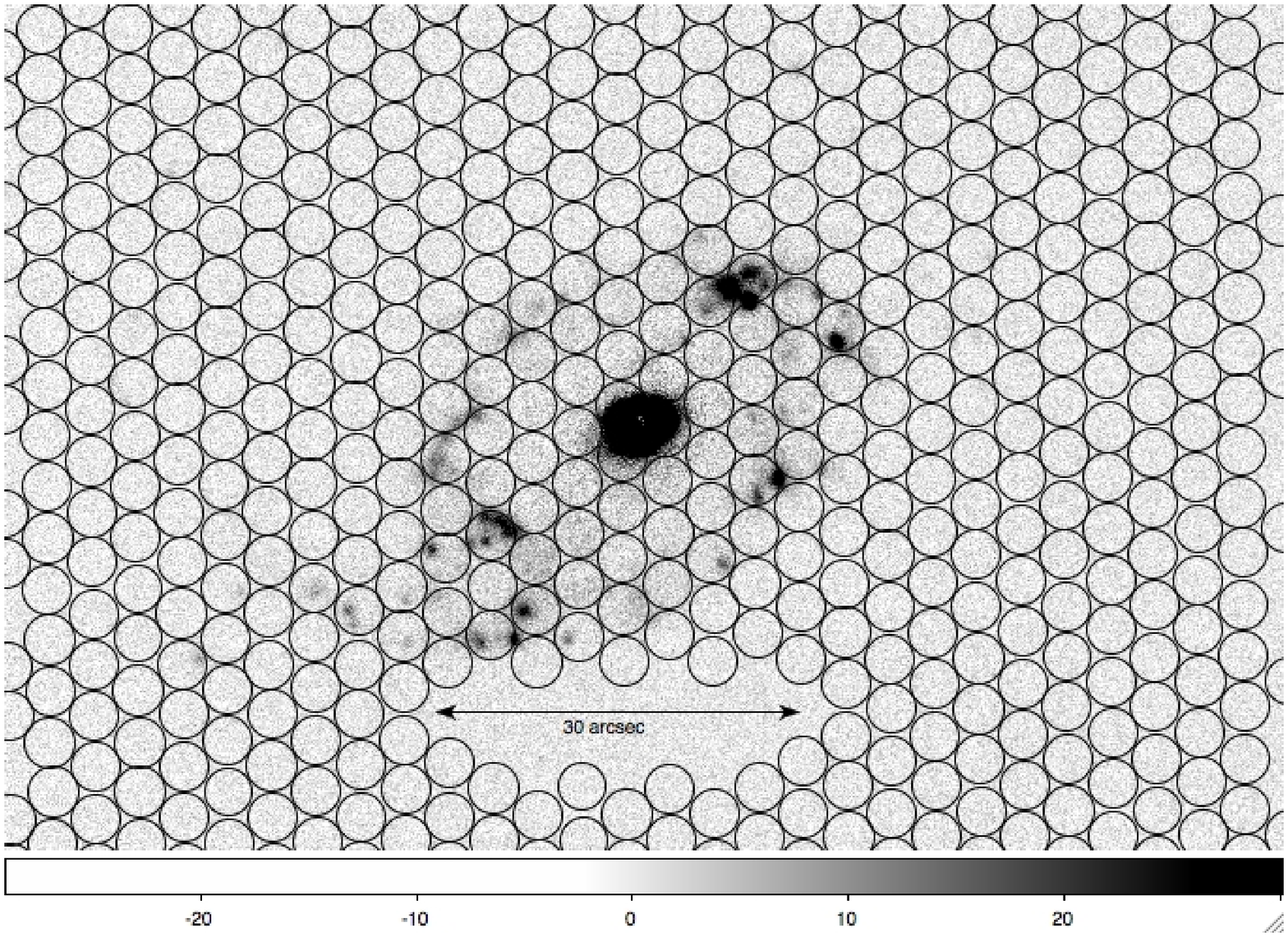}}
\newline
\newline
\newline
\newline
          \subfigure[NGC 7529]{\includegraphics[scale=0.2]{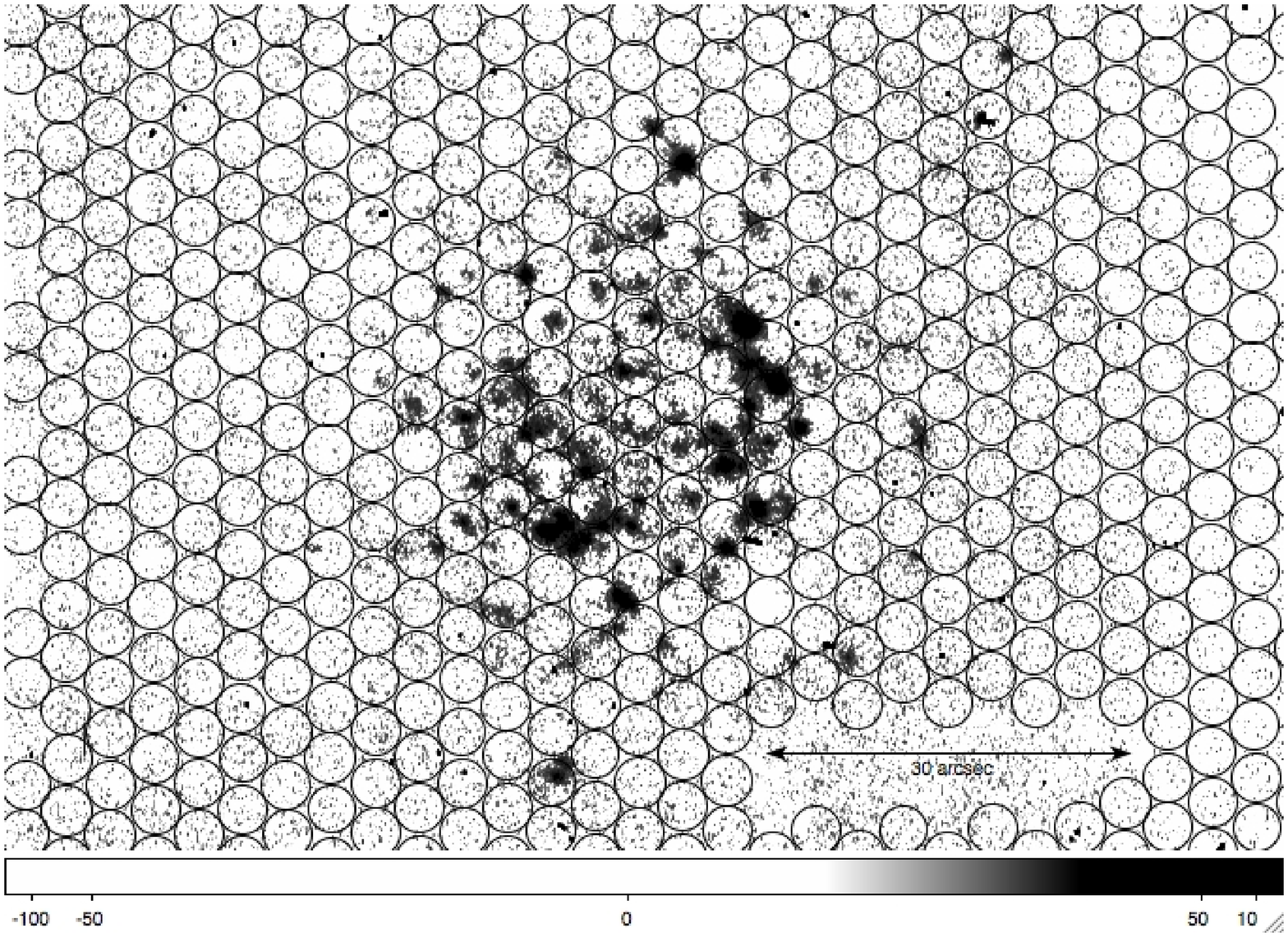}}
          \subfigure[NGC 7591]{\includegraphics[scale=0.2]{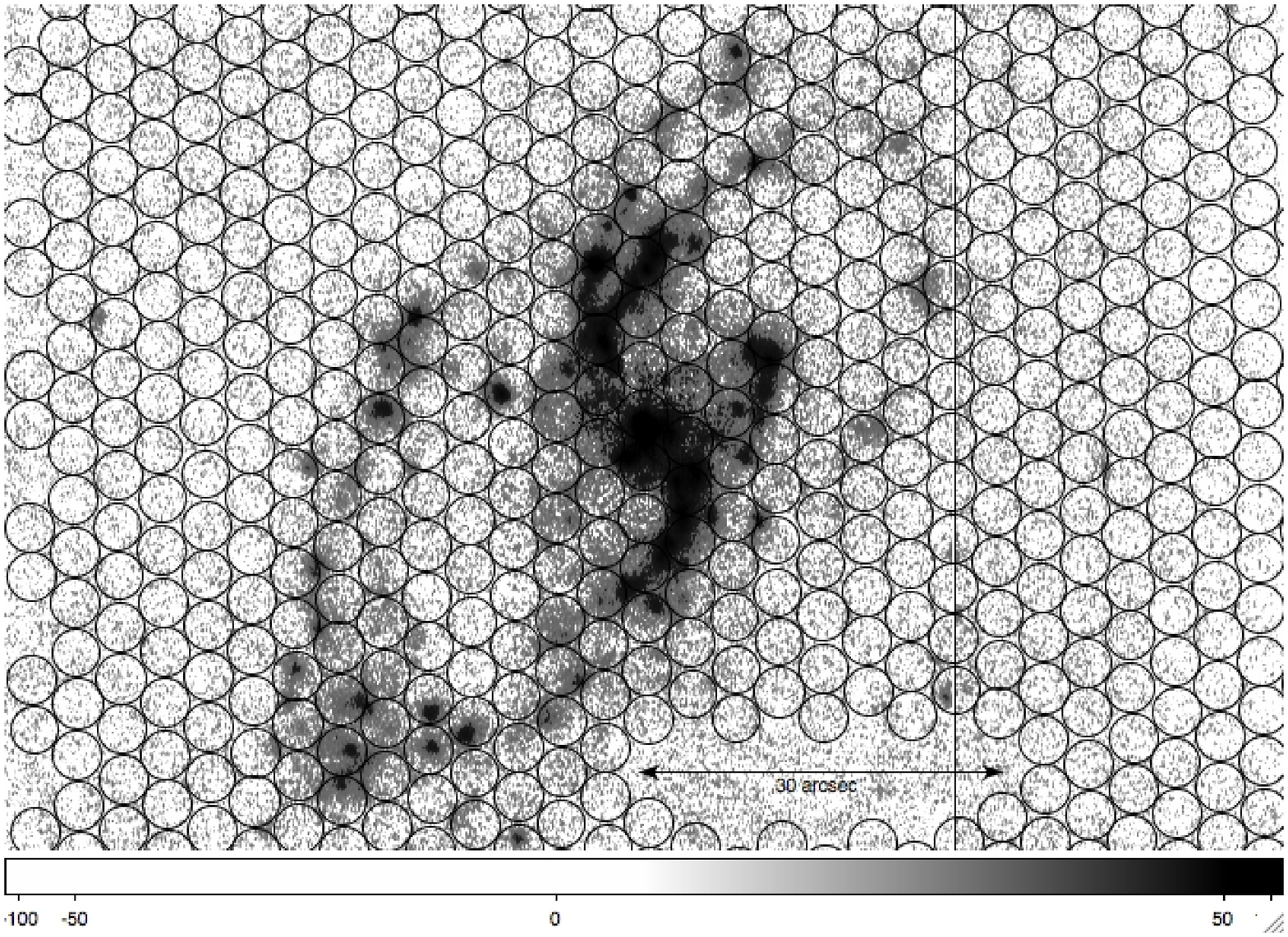}}
          \subfigure[IC 1474]{\includegraphics[scale=0.2]{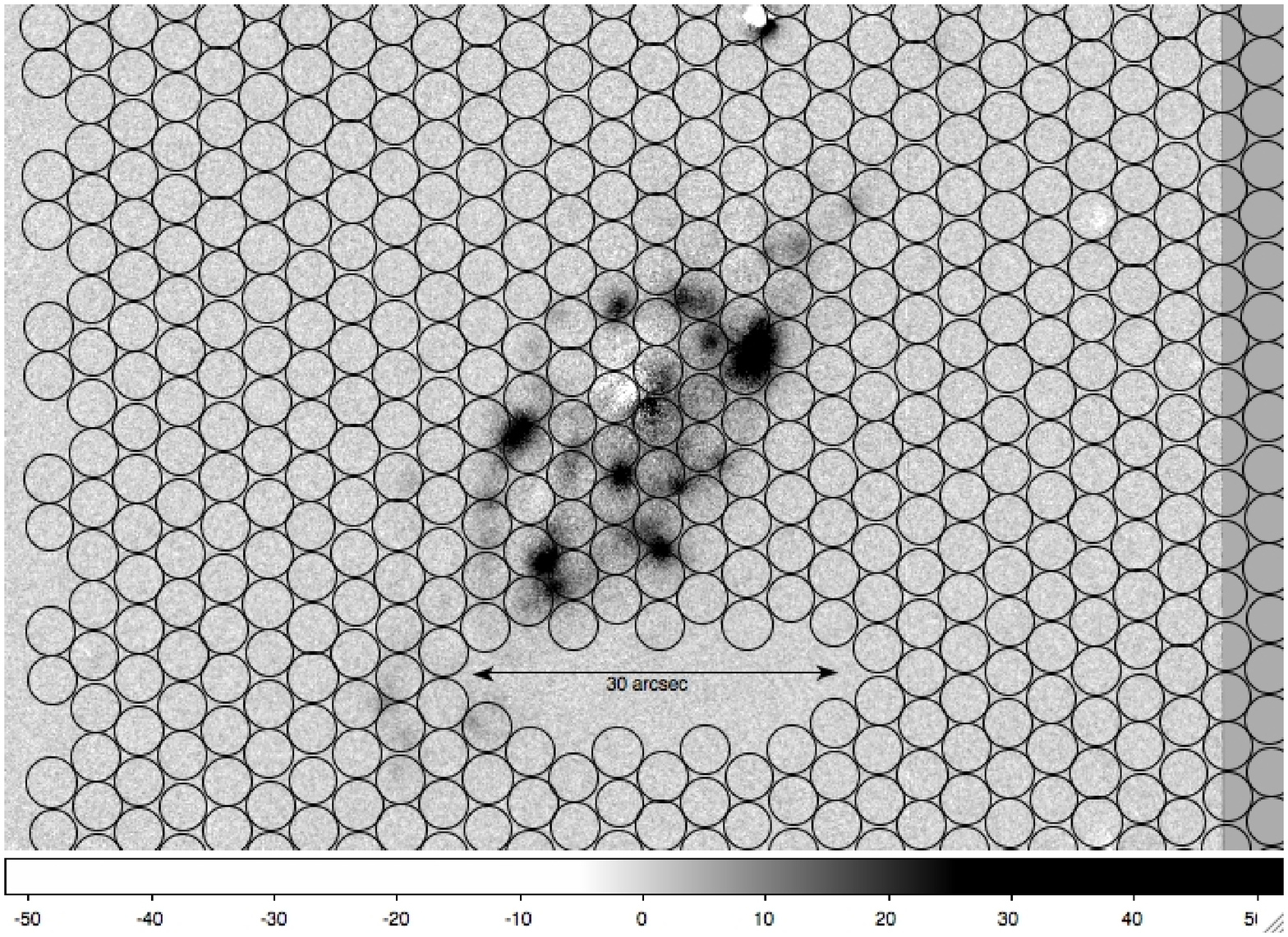}}
	\caption{H $\alpha$ images of Pegasus I cluster spirals included in our sample.  The first row (a-c) shows our hydrogen-deficient targets, while the second row (d-f) presents our gas-normal control set.  The black circles indicate the locations of the 4-arcsecond diameter VIRUS-P fibers, and the arrows are each 30 arcseconds in length.}
	\label{targetfields}
\end{figure}

Our data were obtained during observing runs from 16-20 September 2009 and 14-15 August 2010.  For each galaxy, we observed a three-dither pattern in order to ensure full coverage of the disk with VIRUS-P's fiber field.  Figure \ref{targetfields} shows the narrowband H $\alpha$ images of our targets from Rose et al. (2010) and the fiber maps of the three VIRUS-P dithers for our pointings.  Note the relatively small radii at which H II region emission ceases for our H I deficient sample relative to the control galaxies.  

Wherever possible, we have obtained spectra of our sample using two wavelength settings on the VIRUS-P spectrograph.  The ``blue'' setting covers approximately 3600-5600 \AA, while the ``red'' setting covers approximately 4600-6900 \AA.  The resulting spectral coverage includes emission lines for [O II], [O III], H $\alpha$, H $\beta$, [N II], and [S II].  Furthermore, H $\beta$ is available in both settings as a normalization value.  In each wavelength setup, we exposed for 1 hour on each dither, for a maximum of 6 hours of exposure on each galaxy.  Our targets fit easily on VIRUS-P's 3.5 arcmin$^2$ field of view, so multiple pointings were not required.  For every galaxy in our sample, we have a complete set of dithers on the ``red'' wavelength setting, and we have at least one dither in the ``blue'' setting for every galaxy except IC 1474.  Our treatment for H II regions without blue data will be discussed in subsequent sections.

\section{\bf Data Reduction}
For each pointing on a galaxy, we acquired spectra from 256 fibers.  From these, we have selected those fibers which fall on H II regions using the H $\alpha$ images taken in \citet{rose10}.  As seen in Figure \ref{targetfields}, the 4$\arcsec$ fiber diameter is a good match to the typical angular diameter of H II regions in the Pegasus cluster, so in most cases the spectrum of an individual H II region is contained within a single fiber.  Therefore, the spectra from multiple dithers have not been combined, with the single exception of the H II region labeled (+6.0,-10.2) in IC 1474.

\begin{figure}
  \begin{center}
    \includegraphics[scale=0.6]{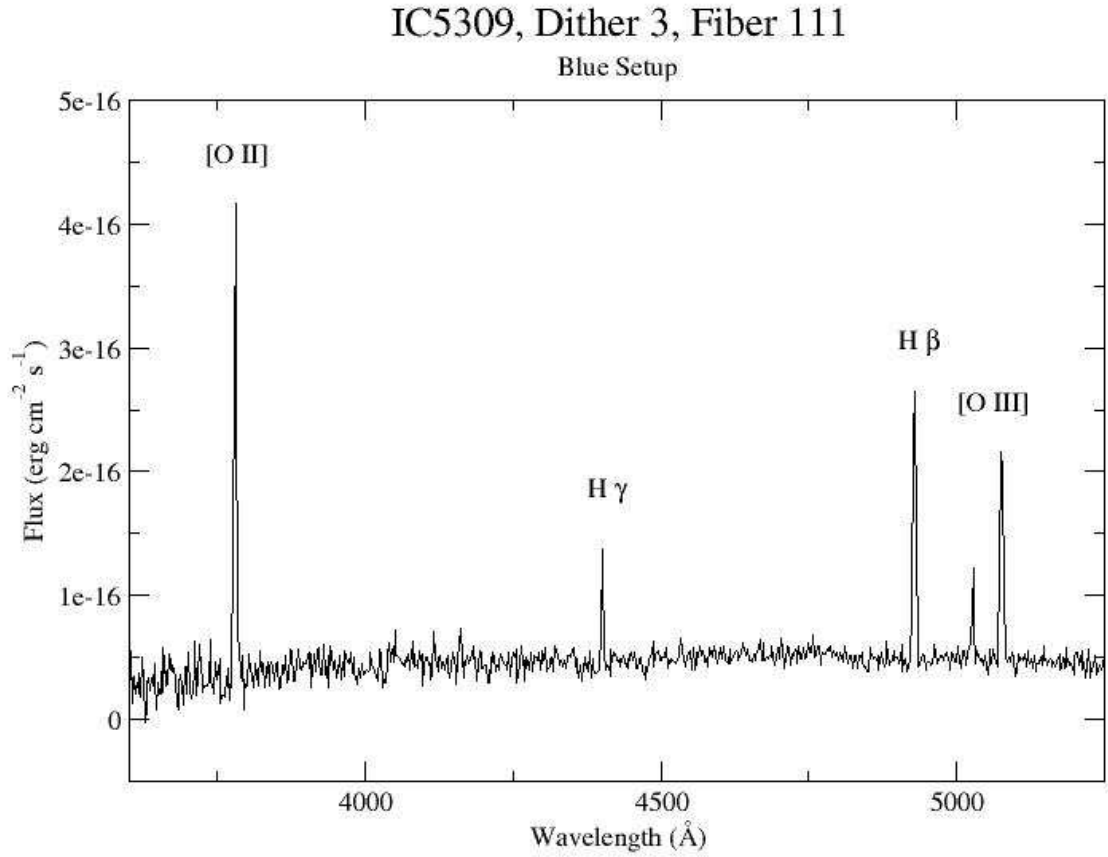}
    \caption{Example H II region spectrum from VIRUS-P.  This fiber was taken on the ``blue'' wavelength setting.  The H II region shown here is from NGC 7529, and is labeled (-1.6, -6.1) in Table 2.}
    \label{examplespec}
    \end{center}
\end{figure}

Figure \ref{examplespec} shows a typical H II region spectrum for a Pegasus spiral.  Basic reduction steps such as bias subtraction, flat fielding, and wavelength calibration were done using the VACCINE software suite for VIRUS-P \citep{hill08}.  Wavelength calibration (accurate to at least 0.5 \AA) is achieved through observation of NeCd (for the ``red'' setting) or HgCd (``blue'' setting) emission lamps, while twilight sky spectra are used to correct uneven pixel response across the CCD.  Each night we also observed at least one white dwarf standard using a six-dither pattern to completely cover the stellar PSF, which we use for flux calibration.  We then measured the emission lines using the splot task in IRAF\footnote[1]{IRAF is distributed by the National Optical Astronomy Observatories, which are operated by the Association of Universities for Research in Astronomy, Inc., under cooperative agreement with the National Science Foundation.}.  Following \citet{skillman96}, the error bars quoted for our line fluxes are obtained by multiplying the RMS of the continuum adjacent to each line by the line width.  Tables \ref{bluefluxtab} (blue) and \ref{redfluxtab} (red) contain the measured emission-line fluxes for our sample, corrected for reddening and stellar absorption, normalized to H $\beta$ = 100.

\subsection{Correction for Balmer Absorption}
Following the method of \citet{skillman96}, among others, we have added a constant 2 $\AA$  EW correction to all Balmer lines in each H II region spectrum to account for underlying stellar absorption.  Our correction follows Equation 3 of \citet{kong02}.  This adjustment eliminates a strong dependence of the H$\alpha$/H$\beta$ ratio on the EW of H $\beta$ apparent in our line fluxes before this correction.

\subsection{Reddening Correction}
For each H II region examined, we determined the reddening coefficient $c$ from the H $\alpha$/H $\beta$ ratio in the ``red'' spectrum.  Fluxes were corrected to the ``case B'' limit for H I recombination lines, namely H $\alpha$/H $\beta$ = 2.86, using the $R = 5.5$ extinction curve, $f(\lambda)$, from \citet{osterbrock06}.  For each H II region, we determine the reddening constant $c = (f_{H \alpha} - f_{H \beta})^{-1}$ log$(I_{H \alpha}/2.86I_{H \beta})$, which we include in Table \ref{redfluxtab}.  Two H II regions have small negative reddening constants, possibly resulting from errors in the line fluxes or in the stellar absorption correction.

\subsection{Abundance Determination}

\begin{figure}
\subfigure[\label{logo}]{\includegraphics[scale=0.5]{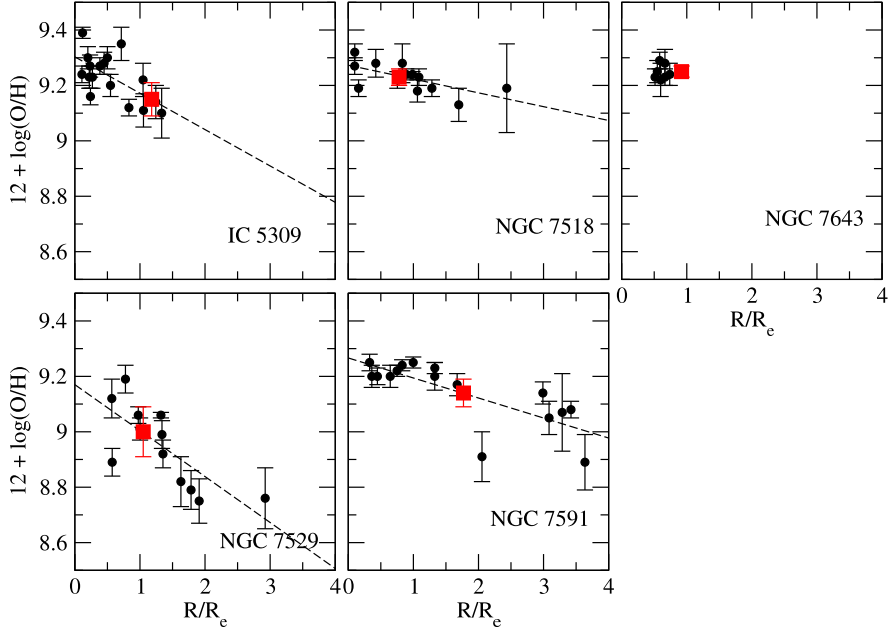}}
\subfigure[\label{o3}]{\includegraphics[scale=0.5]{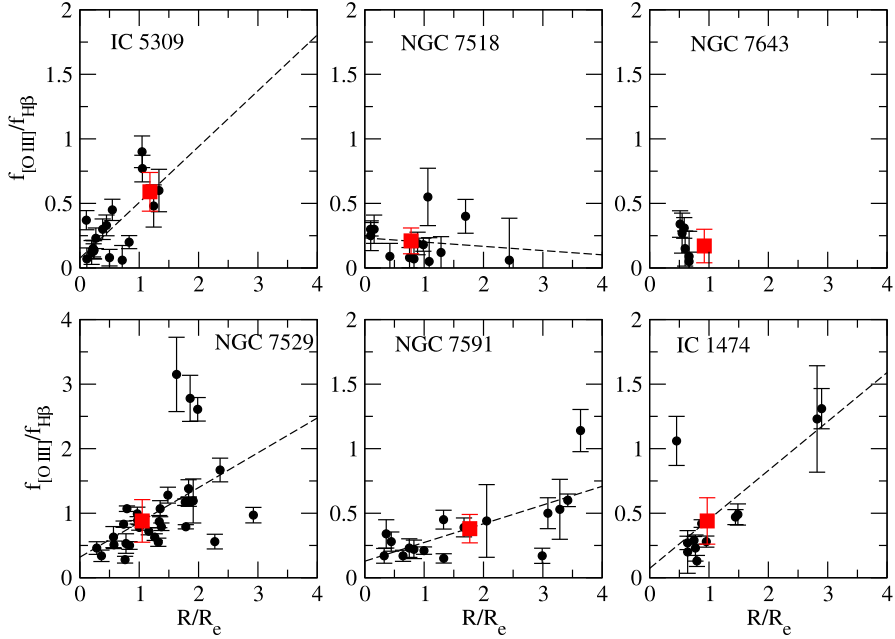}}
\caption{Radial plots of [a] 12 + log(O/H) and [b] $f_{[O III]}/f_{H \beta}$ for H II regions in our sample.  H I deficient galaxies are shown in the top row, while H I normal galaxies are shown on the bottom row.  The dashed lines are our best-fit linear gradients, and the red boxes indicate the mean galactic values.  The mean values are plotted at 0.4 $R_{iso}$, which is where we evaluate the galactic metallicity.}
\label{logovr}
\end{figure}

As with similar studies of extragalactic H II region abundances \citep[e.g.][]{zaritsky94,skillman96}, we use log(O/H) as a proxy for the total heavy-element abundance.  For the individual H II regions in Table \ref{bluefluxtab}, we determine the overall oxygen abundance using the empirical strong-line emission calibration of \citet{zaritsky94}.  We choose to use the oxygen lines from the ``blue'' grating setup for consistency (the [O II] and [O III] lines come from the same spectrum), and so that the weak [O III] lines will fall on the red end of the CCD which, as can be seen from the error bars in Tables \ref{bluefluxtab} and \ref{redfluxtab}, tends to be less noisy.  The \citet{zaritsky94} O/H calibration is based on the quantity $R_{23} \equiv$  ([O II] + [O III])/H $\beta$.  For an illustration of the behavior of $R_{23}$ as a function of log(O/H), see Figure 1 of \citet{mcgaugh91}.   For $\rm log (O/H) > -4.0$, as nebular abundance increases, the greater efficiency of collisional cooling in the fine structure lines lowers the gas temperature, thereby weakening the optical  [O II] and [O III] lines.  In addition, the ionization drops with increasing abundance, because of effects on the temperature and spectrum of the ionizing stars; and this makes the decrease in [O~III] with increasing O/H particularly strong.  While $R_{23}$ is degenerate in log(O/H), the turnaround occurs at very low metallicity (log(O/H) $\sim -4.0$).  Were the oxygen emission decreasing as a result of lower abundance rather than lower temperatures, we would anticipate the [N II] emission, which also traces nebular metallicity, to be low as well.  The [N II] emission lines (Table \ref{redfluxtab}) are sufficiently strong in all H II regions considered here that we can safely assume abundances high enough that $R_{23}$ should decrease monotonically with increasing log(O/H).  In Figure \ref{logo}, we plot 12 + log(O/H) for each H II region versus its galactocentric radius (as measured by $R/R_{e}$).  H~II regions for which we did not obtain blue spectra are omitted due to a lack of [O II] data.

Determining the global nebular abundances for the Pegasus spirals requires some care.  As is evident from Figure \ref{logo}, radial truncation of H II region emission as a result of gas stripping limits our radial coverage of the H I deficient spirals relative to the control sample.  Comparing averages over all H II regions would therefore introduce a bias, given the typical presence of a radial abundance gradient  \citep[see, e.g.][]{kennicutt93,zaritsky94,garnett97}.  A better alternative is to use the metallicity at some characteristic radius.  \citet{zaritsky94} compare several choices for such a fiducial radius, and conclude that 0.4 of the isophotal radius, $R_{iso}$ (the radius for which the R-band surface brightness is equal to 25 magnitudes per square arcsecond), is both reflective of the global O/H content and relatively immune to contamination effects.  Because we have few H~II regions at 0.4 $R_{iso}$, we fit a linear least squares slope for 12 + log(O/H) versus $R/R_{e}$ for all H II regions shown in Figure \ref{logo}.  The mean oxygen abundance for each galaxy is then taken to be the value of this fit at 0.4 $R_{iso}$, with uncertainties derived from the errors of the fitted slopes and intercepts.  For all of our galaxies, we use the values of $R_{e}$ and $R_{iso}$ listed in \citet{rose10}.  In the case of NGC 7643, for which the H II regions span only a small range of radius, we simply take a weighted average over all values of 12 + log(O/H), and adopt the average scatter around the mean as our uncertainty.  As seen in Figure \ref{logovr}, our measured H II regions for NGC 7643 are close enough to $0.4 R_{iso}$, and the fitted gradients are shallow enough that this average should be a reasonable estimate.  In Table \ref{logotab}, we give the resulting galactic mean abundances and gradients for our sample.

As a consistency check, and to include the H II regions for which we do not have [O II] lines, we examine the [O III] lines separately.  Figure \ref{o3} shows the [O III]/H $\beta$ flux for each H II region versus galactocentric radius.  For the sake of uniformity, all flux ratios plotted in Figure \ref{o3} are taken from the ``red'' grating setting.  Recall that the [O II] + [O III] flux increases with decreasing metallicity, hence the inverted slopes between Figures \ref{logo} and \ref{o3}.  Using the same method as described above for the mean abundance, we evaluate the mean [O III]/H $\beta$ for each galaxy, which we include in Table \ref{logotab}.  We plot the mean galactic oxygen measurements against DEF in Figure \ref{avgvdef}.  From inspection of Figures \ref{o3} and \ref{o3vdef}, there is considerably higher scatter and uncertainty for the [O III]-only data.  Furthermore, since the [O III] flux is dependent on the ionization state of the gas as well as the metallicity, we do not include any H II regions without [O II] data in our analysis.

\section{\bf Analysis}

\begin{figure}
\begin{center}
\subfigure[\label{avgvdef_comp}]{\includegraphics[scale=0.3]{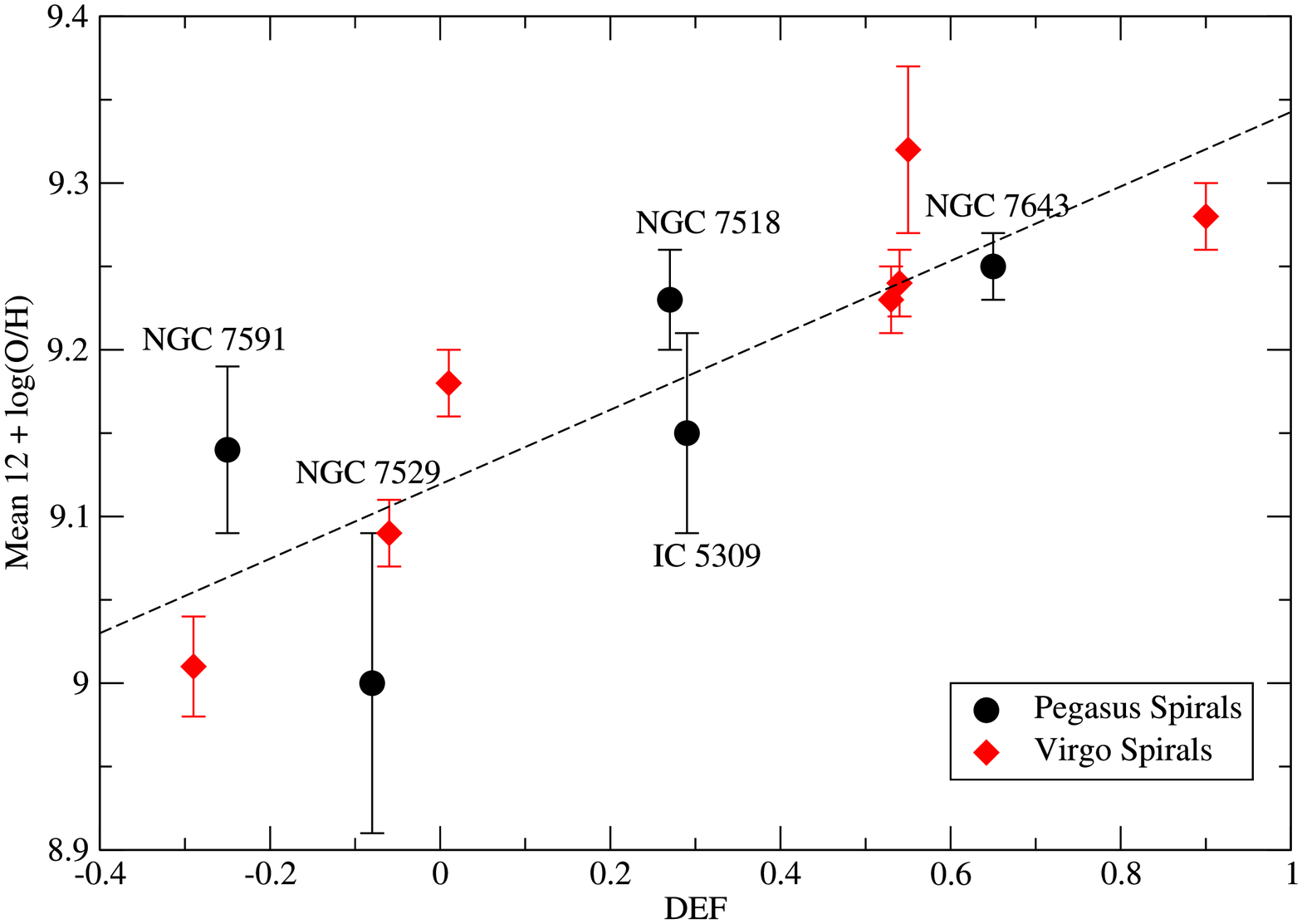}}
\subfigure[\label{o3vdef}]{\includegraphics[scale=0.3]{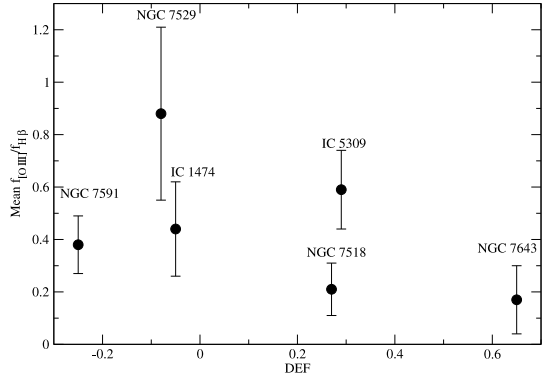}}
\subfigure[\label{avgvdef_field}]{\includegraphics[scale=0.3]{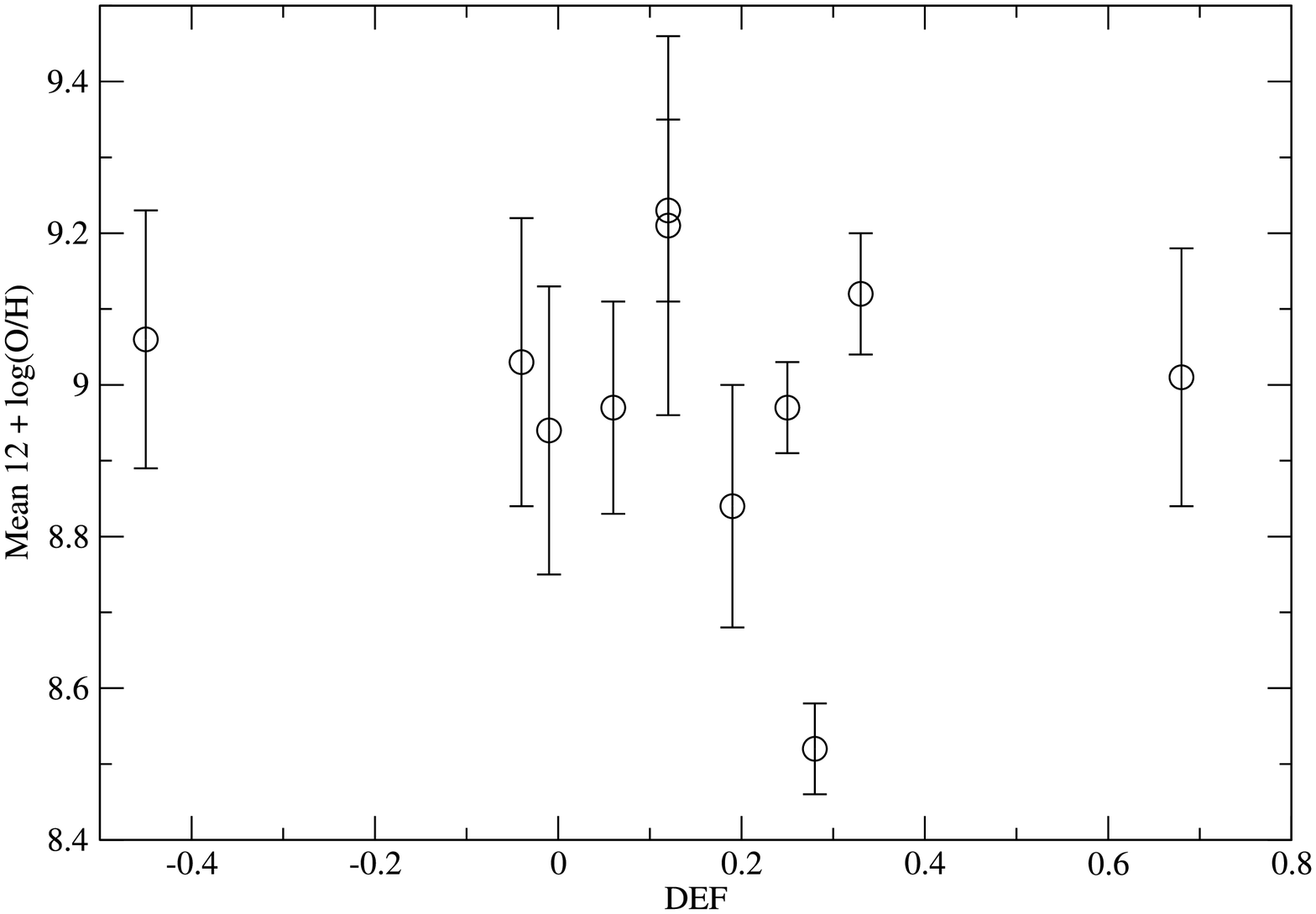}}
\end{center}
\caption{[a]: Mean galactic values of 12 + log(O/H) for Pegasus (black circles) and Virgo (red diamonds) spirals as a function of H I deficiency (DEF).  The dashed line indicates our linear least-squares fit to the data.  [b]:  Mean galactic $f_{[O III]}/f_{H \beta}$ for Pegasus spirals as a function of DEF.   [c]: Mean galactic values of 12 + log(O/H) for unbarred field spirals from \citet{zaritsky94}.}
\label{avgvdef}
\end{figure}

Inspection of Figure \ref{logo} shows considerable overlap in the metallicities of the central H II regions between the gas-deficient and gas-normal galaxies in the Pegasus cluster.  However, Figure \ref{avgvdef_comp} shows a significant trend of increasing mean log(O/H) with DEF.  Allowance for the higher mass of NGC 7591 (see below) strengthens this conclusion.


We see from the mean galactic log(O/H) values that our sample is divided into the expected H I deficient/metal rich and H I normal/metal poor groups for 4 of the 5 objects for which we can determine abundances.  However, NGC 7591 bears discussion as it only marginally conforms.  This object is more than a magnitude brighter than the other galaxies considered, and its circular velocity is the highest of our sample (see Table \ref{targettab}).  \citet{zaritsky94} establish metallicity dependences on $V_C$ and $M_{B}$--both of which are tracers of mass--for spiral galaxies.  Several more recent studies \citep[e.g.][]{tremonti04,mouhcine07,finlator08} have confirmed the observational and theoretical veracity of the so-called ``mass-metallicity relationship'' (MZR).  From Figure 10 of \citet{zaritsky94}, we see that average galactic O/H varies significantly with these properties.  Given the higher mass indicated by $M_B$ and $V_C$, the MZR appears to be a likely reason for the otherwise anomalously high O/H content of NGC 7591 for its value of DEF.  Indeed, similar scatter can be seen for the Virgo cluster, with NGC 4501 (DEF = 0.55) having a higher abundance (log(O/H) = 9.32) than the fainter, more gas-poor NGC 4689 (DEF = 0.90, log(O/H) = 9.28) \citep{skillman96,gavazzi05}.  We therefore conclude that the relatively high log(O/H) observed for NGC 7591 does not contradict the correlation between H I content and nebular abundance observed for the other galaxies.

\begin{figure}
\begin{center}
\subfigure[\label{avgvvc_field}]{\includegraphics[scale=0.3]{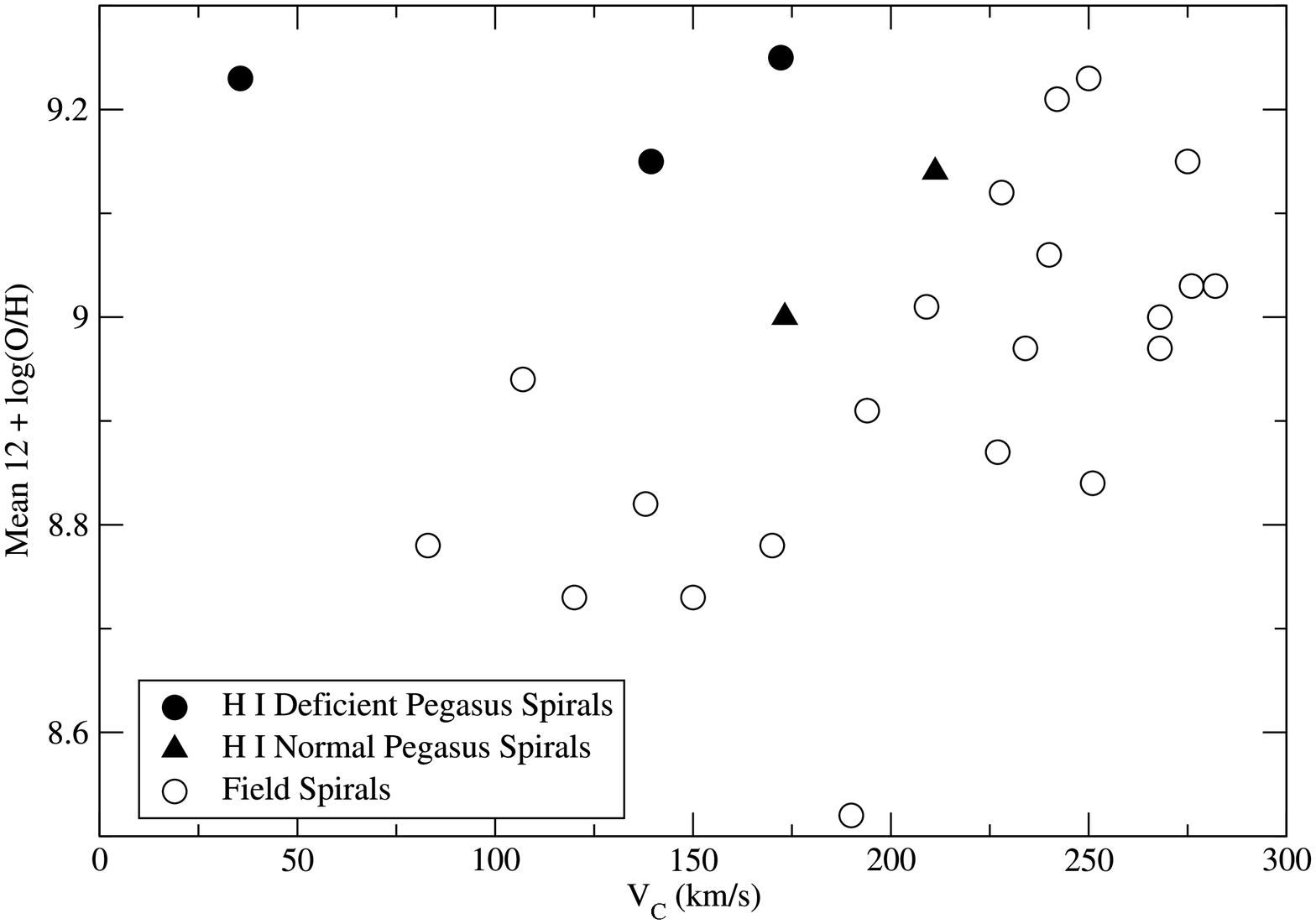}}
\subfigure[\label{avgvmag_field}]{\includegraphics[scale=0.3]{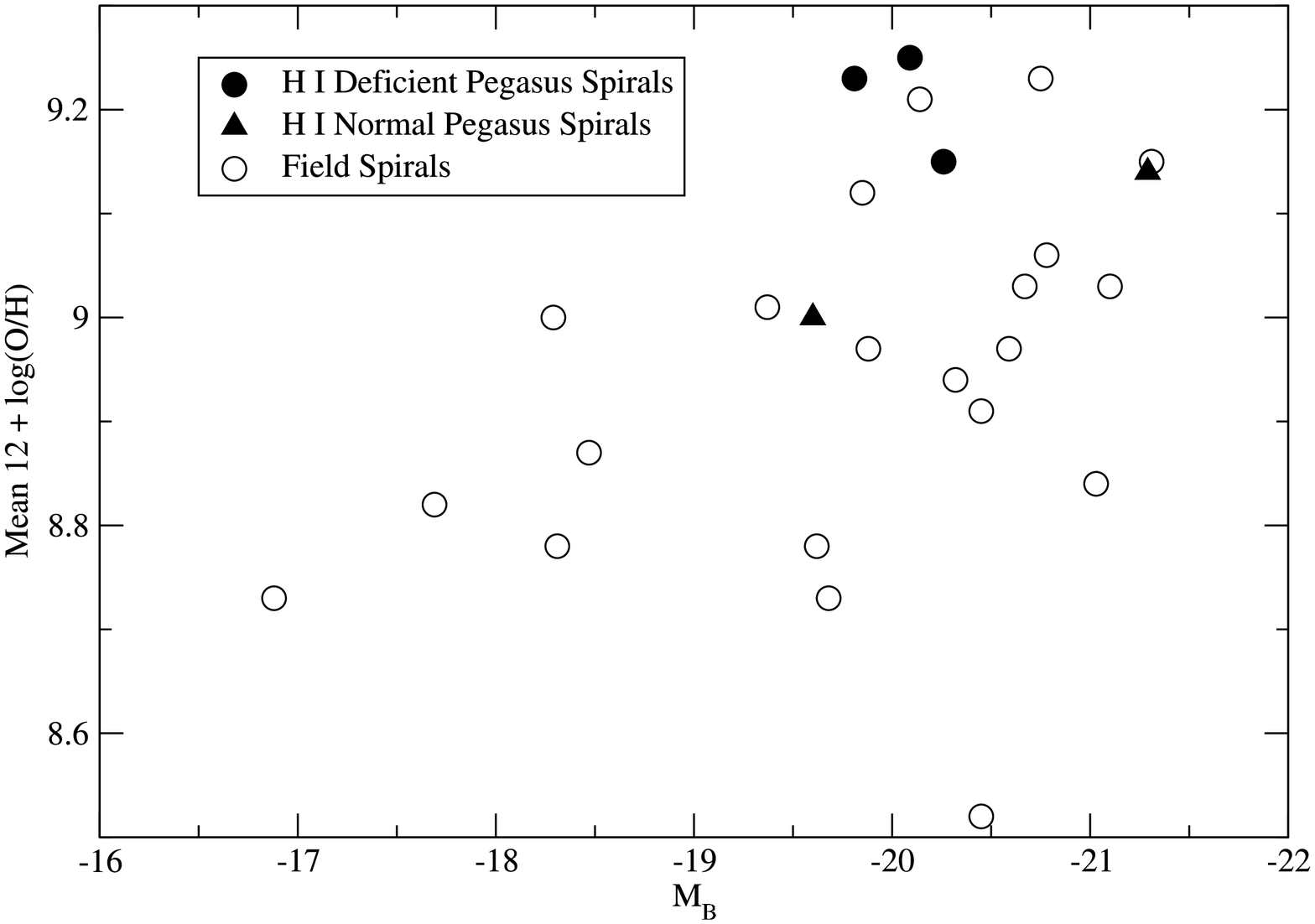}}
\end{center}
\caption{Mean galactic 12 + log(O/H) as a function of [a] inclination-corrected circular velocity and [b] absolute blue magnitude for gas-deficient Pegasus spirals (filled circles), gas-normal Pegaus spirals (filled triangles), and a sample of unbarred field spirals from \citet{zaritsky94} (open circles).}
\label{avg_field}
\end{figure}

It is also interesting to consider the abundances of the Pegasus galaxies in comparison to field galaxies.  Figure \ref{avg_field} plots the mean 12 + log(O/H) for our sample along with a group of unbarred field spirals from \citet{zaritsky94}.  We see that, as a whole, the cluster galaxies all fall towards the higher-metallicity end of the plot, and the H I deficient members are among the most metal rich.  For comparison, Figure 6a of \citet{skillman96} shows the same field galaxies alongside Virgo spirals.  In the case of Virgo, the most H I deficient galaxies are clearly more metal rich than the field, but only slightly more so than our sample.

Taking our H I deficient sample as a whole, we find an average of 9.24 $\pm$ 0.02 for the galactic mean 12 + log(O/H), compared to 9.11 $\pm$ 0.05 for the H I normal controls (excluding IC 1474).  The resultant metallicity offset between H I deficient and H I normal spirals is then $0.13 \pm 0.07$ dex.  While this offset is admittedly marginal at $2 \sigma$, and considerably lower than the 0.3 dex offset claimed for Virgo, a more careful comparison of the two samples shows the metallicity enhancement of the clusters to be more similar than these averages suggest.  

While the environments (IGM density, spiral fraction, velocity dispersion) of the Pegasus cluster differ significantly from Virgo, the individual spirals examined herein represent a very similar population to the Virgo galaxies analyzed in \citet{skillman96}.  As shown in Tables \ref{targettab} and \ref{virgofieldtab}, the two samples cover a similar range in $V_C$ and $M_B$, indicating comparable galactic masses and recent star formation histories.  Furthermore, our targets were selected from the Pegasus surveys of \citet{levy07} and \citet{rose10}, which selected spirals of similar morphologies to the Virgo spirals to enable comparison between the clusters.  We may therefore compare our results to those of \citet{skillman96}, but doing so requires attention to how our characterization of the H I deficiency compares to that used by Skillman et al.  

Interestingly, the 0.13 dex difference between mean (O/H) values for our ``H deficient'' and ``H normal'' groups is approximately the same as the offsets between the hydrogen ``deficient'' and ``intermediate'' and between the ``intermediate'' and ``normal'' groups of Virgo galaxies found in \citet{skillman96}.  A closer examination of their galaxy selection shows that the similarity between these offsets is not coincidental.  \citet{gavazzi05} compute DEF values for Virgo galaxies, allowing us to compare the gas content of the two samples directly.  While the Virgo cluster contains many more galaxies with very high values of DEF (and therefore presumably higher abundances) than Pegasus, only one galaxy (NGC 4689) in the \citet{skillman96} sample has a higher DEF (0.90) than Pegasus' NGC 7643.  We plot the abundances of the seven Virgo spirals from the \citet{skillman96} survey with DEF values measured by \citet[][Table \ref{virgofieldtab}]{gavazzi05} alongside those of Pegasus in Figure \ref{avgvdef_comp}; the two samples have a great deal of overlap in metallicity-DEF space.  Apparently, we are sampling a range of DEF for Pegasus in which our ``gas-normal'' group is analogous to the ``intermediate'' Virgo galaxies, with the ``deficient'' groups being similar for both clusters.  Our metallicity offset is therefore consistent with what we know of the abundance - DEF correlation in Virgo.  For the Virgo spirals plotted in Figure \ref{avgvdef_comp}, the H I deficient galaxies (as defined by having DEF $> 0.3$) have an average 12 + log(O/H) of 9.25 $\pm 0.03$, while the H I normal (DEF $< 0.3$) spirals have an average 12 + log(O/H) = $9.11 \pm 0.06$.  This offset, $0.14 \pm 0.09$, is neither larger nor more significant than the offset for our Pegasus sample.  Evidently, the larger metallicity offset observed for Virgo is not a result of sampling more stripped, metal-rich galaxies, but from choosing a more remote, gas-rich control sample.  Figure 1 of \citet{skillman96} shows that two of the three H I normal Virgo galaxies examined--NGC 4651 and NGC 4713--are so far from the cluster center as to essentially be field galaxies.  Indeed, neither of these objects appears in the Virgo Cluster Catalog \citep[VCC,][]{binggeli85}, hence their absence in the \citet{gavazzi05} H I survey.  As expected based on the observed correlation, they display very low oxygen content (12 + log(O/H) $<$ 9.00), contributing to the 0.3 dex metallicity offset for Virgo.  We conclude, then, that the process of nebular metallicity enhancement observed in the Virgo cluster has occurred to a similar degree in Pegasus at fixed DEF.  

When evaluating galactic H I deficiency quantitatively with DEF rather than the more qualitative groupings of \citet{skillman96}, it becomes apparent that evaluating the influence of the cluster environment on galactic metallicity with an offset between hydrogen-poor and hydrogen-normal groups can be misleading.  As each sample is likely to have a different range of DEF values, as seen for our sample compared to Virgo, such a bimodal separation leads to ambiguous conclusions.  A better solution is to examine the correlation between log(O/H) and DEF, which more accurately describes the continuous progression towards higher metallicities with increasing H I deficiency.  From Figure \ref{avgvdef_comp}, we see a strong correlation in log(O/H) versus DEF for the combined Pegasus-Virgo sample.  Performing a linear least-squares fit to the trend, we derive the relation $12 + \log $(O/H)$ = 9.120 + 0.223 \times$DEF, with uncertainties of 0.045 dex$_{(O/H)}$/dex$_{DEF}$ on the slope and 0.02 dex$_{(O/H)}$ on the intercept.  The resulting Pearson correlation coefficient to the data is 0.84.  Thus, while the metallicity offsets of the samples above and below DEF = 0.3 are only significant at the $\sim 2 \sigma$ level, we find a slope in metallicity-DEF space that approaches a $5 \sigma$ confidence level.  

While we see that H I deficiency affects the overall nebular metallicity of cluster spirals, it is important to disentangle this process from the secular effects of galaxy mass.  In order to examine the metallicity-DEF correlation independently of the mass-metallicity relationship, we consider the differential (O/H) offset between a galaxy and expectation values based on its $V_C$ and $M_B$.  We adopt $V_C$ as our primary tracer of a galaxy's mass since, as mentioned in \citet{zaritsky94}, it is distance-independent, unbiased by recent star formation, and more tightly correlated with galactic metallicity.  However, we include our analysis in terms of $M_B$ for the sake of completeness.  Fitting a linear trend to the abundance versus circular velocity plot for the \citet{zaritsky94} field sample shown in Figure \ref{avgvvc_field}, we derive a galaxy's expected oxygen abundance as

$12 + \log$(O/H)$ = 8.57+0.356 \times V_C/(200~\mathrm{km/s})$.

Similarly, from Figure \ref{avgvmag_field}, we derive

$12 + \log$(O/H)$ = 8.95 - 0.0594 \times (M_B + 20)$

\begin{figure}
\begin{center}
\subfigure[\label{vcorr}]{\includegraphics[scale=0.25]{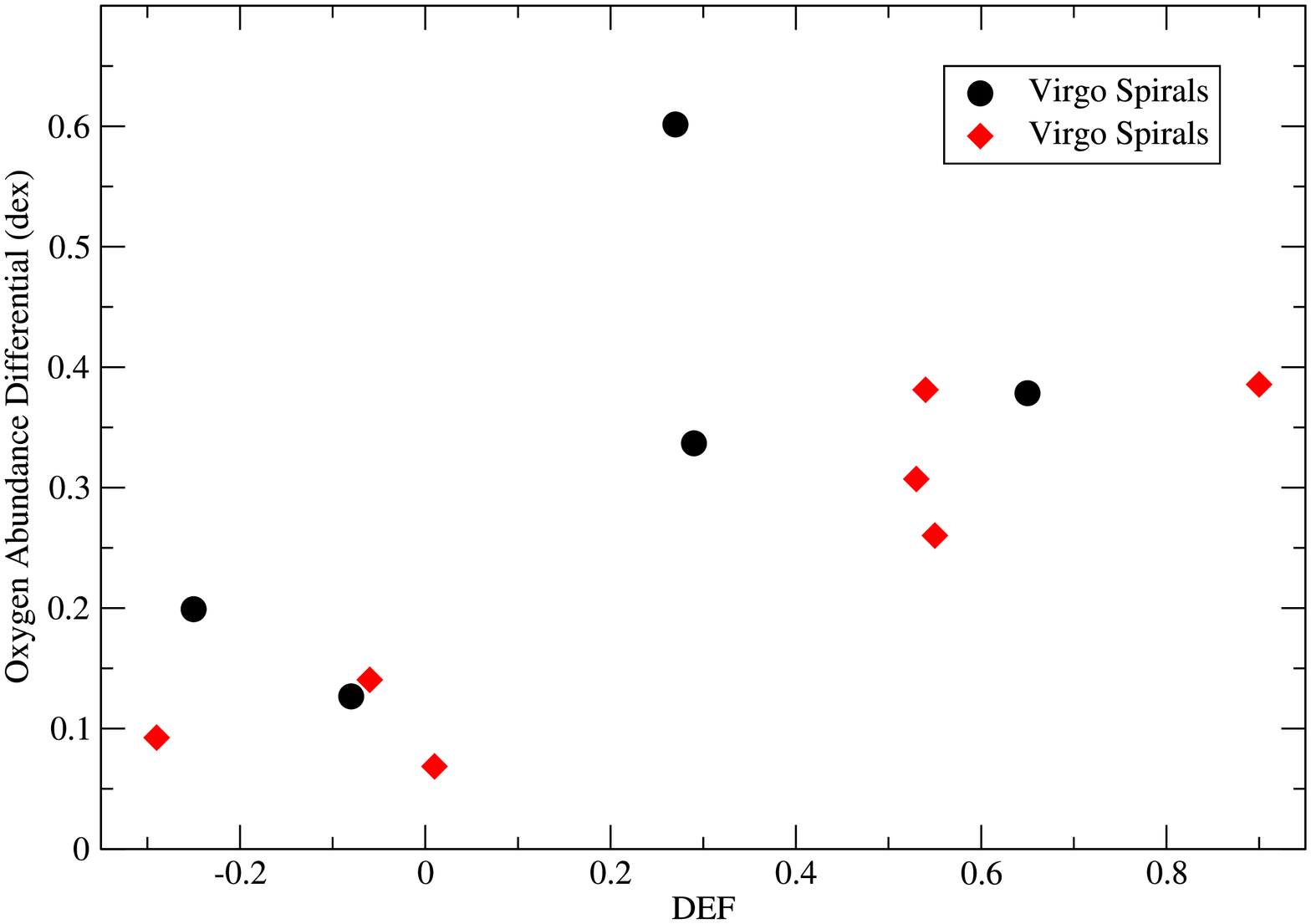}}
\subfigure[\label{vcorr_field}]{\includegraphics[scale=0.25]{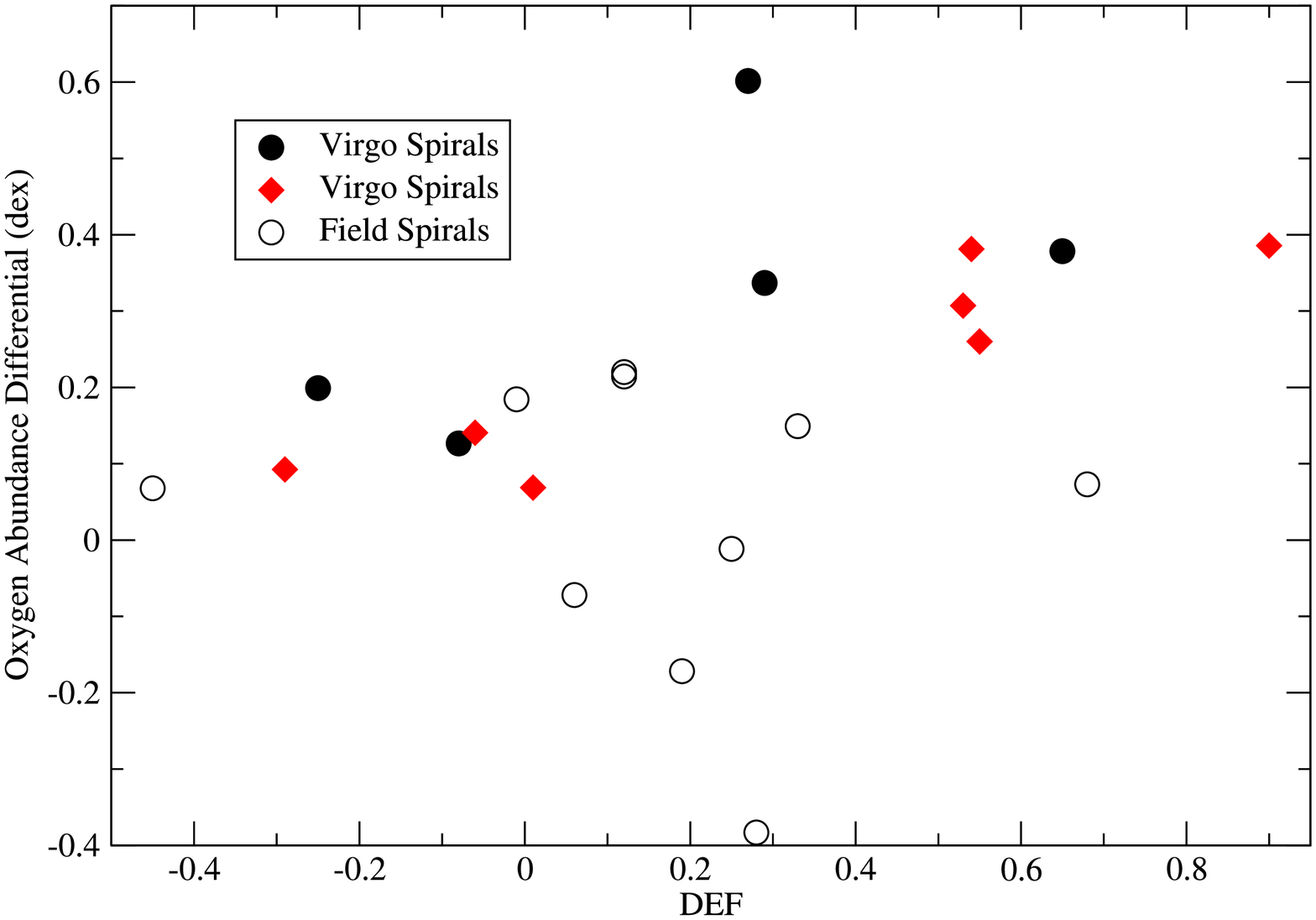}}
\subfigure[\label{mcorr}]{\includegraphics[scale=0.25]{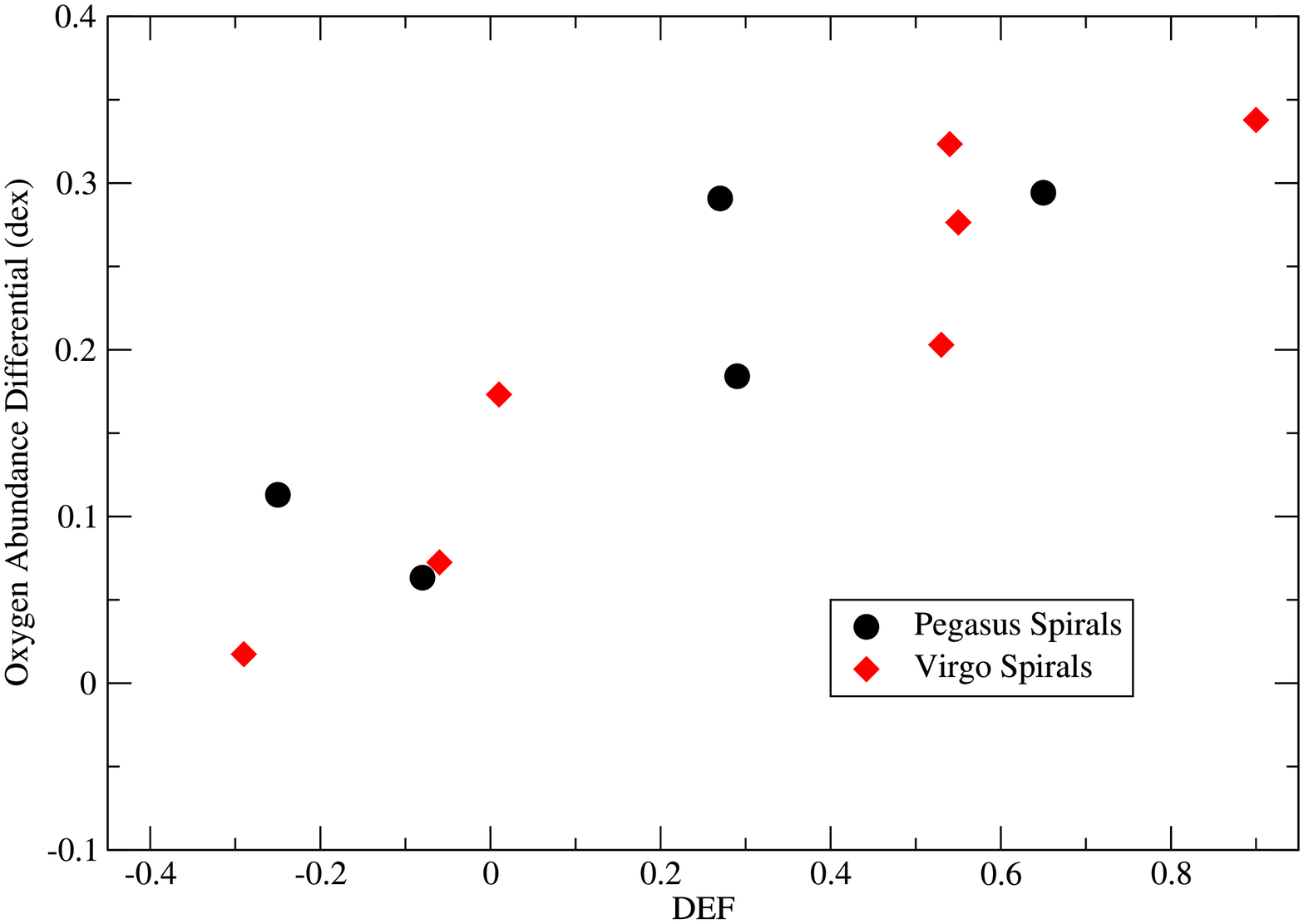}}
\subfigure[\label{mcorr_field}]{\includegraphics[scale=0.25]{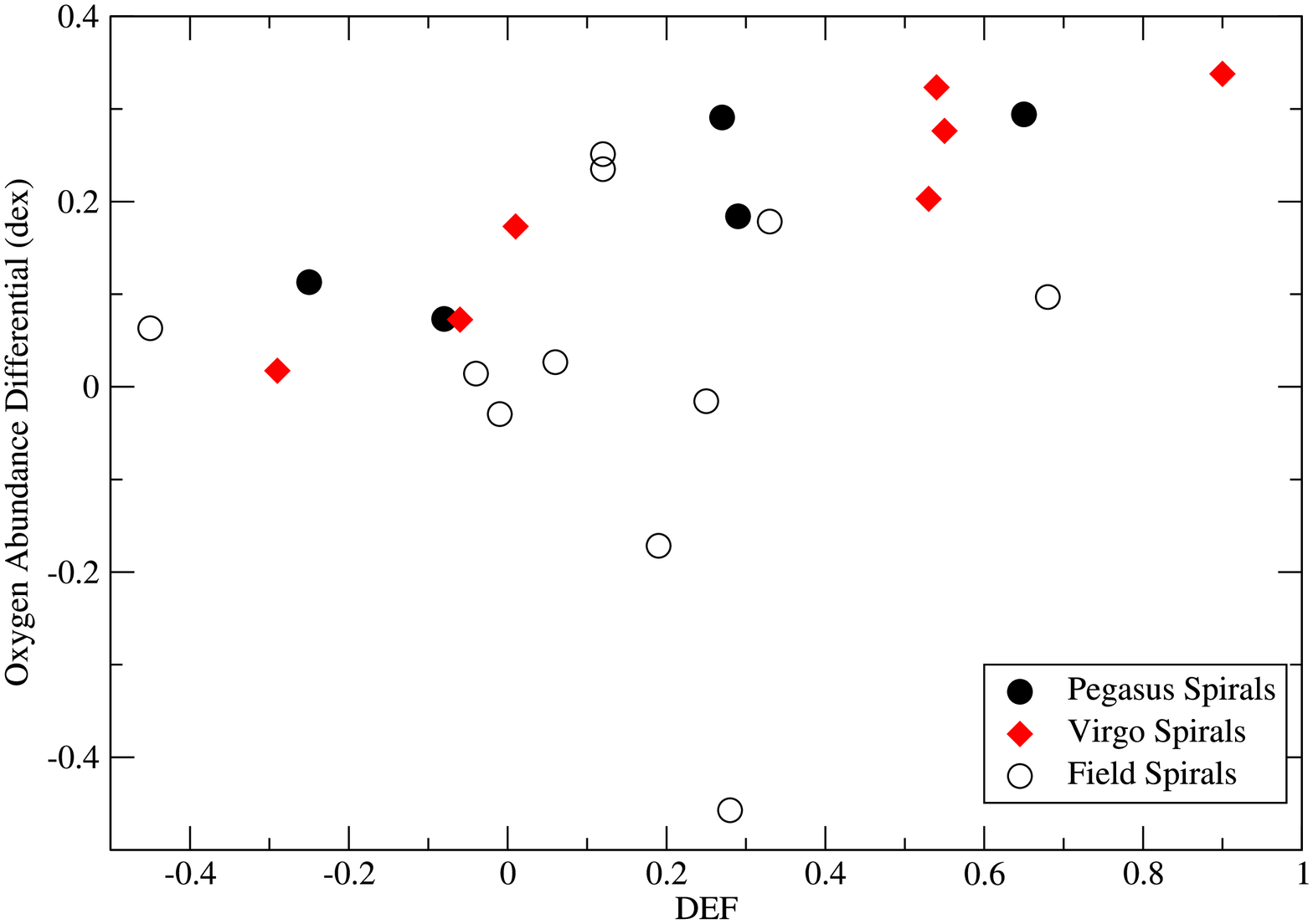}}
\end{center}
\caption{[a,c]: log(O/H) differential (measured - expected) versus DEF for Pegasus (black circles) and Virgo (red diamonds) spirals.  Expectation values are based on the log(O/H)-$V_C$ (a) and log(O/H)-$M_B$ (c) correlations presented in \citet{zaritsky94}.  [b,d]: Same as [a,c], but with the addition of the Zaritsky field spirals.}
\label{avgvdef_corr}
\end{figure}

Figure \ref{avgvdef_corr} shows the offsets in log(O/H) (measured - expected) versus DEF for the galaxies presented in Figure \ref{avgvdef}.  The comparison to expectation values effectively removes the scatter introduced by the MZR, and the resulting correlation is obvious.  For the combined set of Pegasus and Virgo spirals (excluding NGC 7518, which has a $V_C$ of 36 km/s, far lower than any of the other galaxies examined here), the Pearson correlation coefficient reaches 0.86 after removal of the (O/H)-$V_C$ trend.  If we instead remove the (O/H)-$M_B$ trend, the correlation coefficient for the two samples is 0.90.  From this analysis, we conclude that, as expected, at least some of the scatter around the DEF-(O/H) correlation in Figure \ref{avgvdef_comp} is due to the MZR.

\begin{figure}
\begin{center}
\includegraphics[scale=0.6]{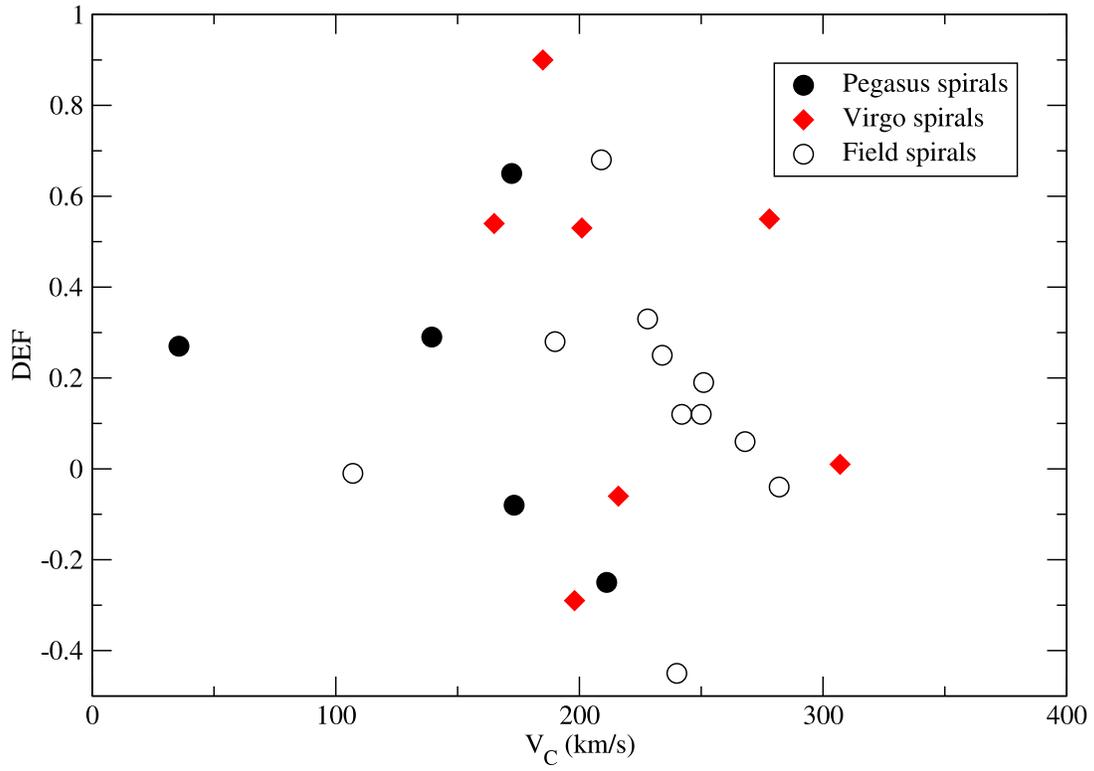}
\end{center}
\caption{H I deficiency parameter DEF as a function of $V_C$ for Pegasus (filled circles), Virgo (diamonds), and field (open circles) spirals.}
\label{defvvc}
\end{figure}

It is important to note that while we see effects of the MZR in our sample, the observed metallicity offsets are not primarily driven by galaxy mass.  This would be particularly likely if the H I deficient galaxies were systematically more massive than the H I normal sample.  If this were the case, we would expect to see a correlation between DEF and $V_C$.  To test this possibility, we plot DEF versus $V_C$ in Figure \ref{defvvc}.  We observe no correlation for Pegasus or Virgo spirals, thus ruling out a mass offset between hydrogen normal/poor spirals.  We can therefore conclude that H I deficiency is driving metallicity augmentation independently of galaxy mass.

The observed correlation between heavy element content and DEF observed for Pegasus and Virgo galaxies might suggest that these objects' metallicity offsets are caused entirely by H I deficiency, and are independent of cluster membership.  However, a comparison to field spirals refutes that notion.  In Figures \ref{avgvdef_field} and \ref{avgvdef_corr}, we have plotted the mean 12 + log(O/H) against DEF for field galaxies from \citet{zaritsky94}.  Our values of DEF for the field galaxies--which we include in Table \ref{virgofieldtab}--are adopted from \citet{fumagalli09}.  We see that, unlike for the cluster spirals, the abundances for field galaxies are completely uncorrelated with DEF. The Pearson correlation test confirms what visual inspection suggests; the (O/H)-DEF correlation coefficient for the field sample is -0.16.  While the measured metallicities for the field sample have a lower precision than our sample or the Virgo sample, it appears very tentatively that heavy element content is only dependent on H I deficiency if a galaxy has lost its H I through cluster-driven mechanisms, as opposed to field objects that have always been gas-poor.  However, further study of both field and cluster spirals will be required to properly evaluate how the (O/H)-DEF correlation changes for different environments.  In particular, it will be essential to obtain abundances for a larger sample of field spirals with DEF $>$ 0.3 to facilitate a more meaningful comparison to Pegasus and Virgo.

\section{\bf Discussion}

The physical cause of the observed metallicity offset for cluster galaxies has been the subject of some debate.  The typical model for gas-phase chemical evolution in an isolated galaxy is a variant of the ``simple model'' \citep[e.g.][]{pagel75}.  The simple model treats a galaxy as a ``closed box'' of hydrogen gas, which grows progressively more metal-rich through stellar recycling according to $Z = -y$ ln$\mu$, where $Z$ is the total metallicity, $\mu$ is the gas fraction, and $y$ is the yield of metals from massive stars.  This model is easily modified to account for the infall of primordial gas.  As mentioned before,  the inflow of unprocessed gas should be terminated when a galaxy falls into the hot ICM, resulting in a metal enrichment pattern that more closely parallels the pure simple model \citep[as in][for example]{henry94}.  \citet{skillman96} explore the possibility that infall cutoff explains the enhanced metal abundance of the core Virgo spirals.  Their chemical evolution models with and without infall demonstrate that infall suppression can create a metallicity increase of ~0.15 dex--a significant fraction of the Virgo offset, but not sufficient to account for the entire effect.  In the case of Pegasus, though, 0.15 dex is adequate to produce the observed enrichment.  One possibility, then, is that as a galaxy first falls into a cluster environment, infall cutoff is the primary external driver of chemical evolution, producing a metallicity transition akin to the H I normal-intermediate shift seen for Virgo.  As galaxies continue to fall through the ICM and experience increasingly more H I stripping, additional processes associated with the more extreme H I deficiencies observed become important, resulting in the elevated abundances seen for the ``deficient'' spirals.

A second mechanism proposed to drive metal enhancement requires the elimination of radial inflows of metal-poor gas.  As described in \citet{shields91}, if the galactic H I disk is truncated at a radius interior to the stellar disk, then the inward transport of low-metallicity gas from the exterior of the galaxy is inhibited, preventing the dilution of metals in the galactic interior.  Without this dilution, the abundance gradients typically seen for field galaxies tend to flatten out, leading to high characteristic metallicities.  The H $\alpha$ emission of Pegasus spirals is clearly truncated inside the stellar disks for H I deficient members \citep{rose10}.  Furthermore, the flattened metallicity gradients predicted for inflow cutoff by \citet{shields91} are precisely what we observe for our targets.  If the radial inflow effect is comparably important as infall cutoff for Virgo, the same may be true for the Pegasus cluster. 

Our results add to a growing number of studies demonstrating evidence for environmental influences on galactic chemical evolution.  \citet{tremonti04} use strong-line calibrations for integrated spectra \citep[which are reflective of a galaxy's mean metallicity, see][]{moustakas06} of SDSS galaxies to determine 12 + log(O/H).  \citet{cooper08} analyze these abundances, finding a strong dependence on local galaxy density (as defined by nearest-neighbor analysis), particularly for objects in group/cluster environments.  Similarly, \citet{ellison09} use nebular abundances from SDSS galaxies in \citet{kewley02} to demonstrate a metallicity-galaxy density relationship, although they claim the effect depends only on local density, not cluster membership.  However, using the local galaxy density alone has its limits, as very close galaxy pairs induce inflow of metal-poor gas through tidal interactions, leading to \emph{lower} characteristic abundances than isolated field galaxies \citep{kewley06,ellison08}.  These results bear particular significance for NGC 7643, which is within $\sim 200$ kpc of UGC 12562 \citep{levy07}.  While this separation is too large by an order of magnitude for tidal interactions to lower the metallicity of NGC 7643 \citep{kewley06}, it is possible that ``galaxy harassment'' has contributed to that galaxy's high DEF, and therefore elevated (O/H) abundance.  

The metallicity dependence on gas content is confirmed by \citet{zhang09}, who cross-correlate the galaxies in \citet{tremonti04} with H I data in the HyperLeda catalog, and find that gas-poor galaxies tend to be metal rich.  While our study rejects a dependence on H I content for the metallicities of field galaxies, we note that the \citet{zhang09} galaxy selection does not account for cluster membership, so their observed correlation may be driven by galaxies in clusters.

The importance of environment for galactic metallicity has not been unanimously accepted.  \citet{mouhcine07} examine the \citet{tremonti04} data and find a metallicity correlation with local density that is much weaker than those claimed in \citet{cooper08} and \citet{ellison09}.  \citet{mouhcine07} concede, however, that local environment can drive chemical evolution at least in the cores of dense clusters.  In a study more similar to our own, \citet{petropoulo11} compute oxygen content from spatially-resolved long-slit spectroscopy of star-forming galaxies in the Hercules cluster.  While they do observe a local density dependence for the metallicity of dwarf galaxies, they claim no environmental effect for major spiral galaxies.  We note two caveats here, though.  First, they do not quantify hydrogen deficiency, making comparisons to Pegasus or Virgo difficult.  Additionaly, they have much lower spatial resolution than we have for Pegasus, and they treat the cores and disks of giant spirals as separate objects, making them insensitive to the kind of metallicity offset we extract from the galactic abundance gradients.  We therefore do not consider the \citet{petropoulo11} study to be contradictory to ours.  Overall, the evidence indicates that dense environments like those found in galaxy clusters have a significant influence on the abundance patterns of their member galaxies.

As with Virgo, the heavy element abundance differential for Pegasus is considerably larger than the aforementioned statistical samples of SDSS galaxies.  The combined set of data for these two clusters appears to explain why.  It seems that, while considering cluster/group membership alone will reveal a small metallicity offset, the primary importance of the cluster is to drive H I stripping and infall cutoff, which ultimately leads to proportionally higher heavy element content.  To date, \citet{skillman96} and this study are the only analyses to scrutinize the complementary processes of H I removal and metallicity enhancement as codependent effects of cluster-driven galaxy evolution.  The additions of NGC 7518 and IC 5309 to the metallicity-DEF relationship (Figure \ref{avgvdef_comp}) are particularly interesting, as they fill a previously unsampled intermediate range in DEF, showing the correlation to be continuous rather than bimodal.  We emphasize that future metallicity surveys of additional clusters should pay careful attention to H I content, as we believe it to be essential to understanding the full influence of a cluster on the properties of its member galaxies.

\section{\bf Summary}
We have presented integral-field spectroscopy of six galaxies in the Pegasus I cluster, a young, low-density cluster.  We analyze the spectra of H~II regions in these galaxies with the aid of a calibration of the [O~II] and [O~III] emission-line intensities to determine gas-phase heavy element abundance.  Combining these results with published H~I data, we examine the abundances as a function of H~I deficiency, the possible result of ISM-ICM interactions.  When we account for the effects of the galactic mass-metallicity relationship, we find that oxygen abundance correlates well with the hydrogen deficiency parameter DEF for the Pegasus galaxies. The hydrogen-deficient members of our sample show, on average, higher values of log(O/H) at the 0.15 dex level, which is consistent with Virgo spirals of similar gas deficiency.   Our results agree qualitatively with a number of recent publications indicating intimate connections between a galaxy's heavy element content, hydrogen deficiency, and the density of its local environment.

\begin{acknowledgements}
We thank Sara Ellison, Rob Kennicutt, Josh Adams and Evan Skillman for helpful discussions.  G.S. gratefully acknowledges the support of the Jane and Roland Blumberg Centennial Professorship in Astronomy.
\end{acknowledgements}

\clearpage

\clearpage

\begin{table}

\begin{center}

{Target Galaxies}

\begin{tabular}{l l l l l l l }

	\hline\hline \\
	Galaxy & R.A. & Decl. & $V_{C}$\tablenotemark{2} (km/s) & $M_{B}$\tablenotemark{2,3} & $R_{e}$ & $R_{iso}$ \\ \hline
	\\
	IC 1474 & 23:12:51.2 & +05:48:23 & 138.3 & -19.72 & 14. & 34. \\
	NGC 7518 & 23:13:12.8 & +06:19:18 & 35.6 & -19.81 & 22. & 43. \\
	NGC 7529 & 23:14:03.2 & +08:59:33 & 173.2 & -19.60 & 11. & 29. \\
	NGC 7591 & 23:18:16.2 & +06:35:09 & 211.2 & -21.29 & 14. & 62. \\
	IC 5309 & 23:19:11.7 & +08:06:34 & 139.4 & -20.26 & 19. & 56. \\
	NGC 7643 & 23:22:50.4 & +11:59:20 & 172.2 & -20.09 & 20. & 43. \\
	\hline

\end{tabular}
\caption{Target galaxies, their coordinates (J2000.0), inclination-corrected circular velocities, absolute B magnitudes, effective radii ($R_{e}$), and isophotal radii ($R_{iso}$).  Radii are given in arcseconds.}
\label{targettab}
\end{center}

\end{table}

\tablenotetext{2}{We acknowledge the usage of the HyperLeda database (http://leda.univ-lyon1.fr).}
\tablenotetext{3}{This research has made use of the NASA/IPAC Extragalactic Database (NED) which is operated by the Jet Propulsion Laboratory, California Institute of Technology, under contract with the National Aeronautics and Space Administration.}
\tablenotetext{4}{As defined by \citet{skillman96}}
\tablenotetext{5}{Virgo cluster DEF measurments from \citet{gavazzi05}.  Field galaxy DEF measurements from \citet{fumagalli09}.}
\tablenotetext{6}{Virgo cluster abundances from \citet{skillman96}.  Field galaxy abundances from \citet{skillman96}.}

\clearpage

\begin{center}

\tablecaption{Corrected emission line fluxes for H II region spectra using the VIRUS-P blue setup.}
\label{bluefluxtab}	
\tablefirsthead{\hline
H II region position &  [O II]  &  H $\gamma$ &  H$\beta$  &  [O III]  &  [O III]  &  $R/R_{e}$  \\ 
(arcsec from center) & $\lambda3726 + \lambda3729$ & & & $\lambda4959$ & $\lambda5007$ & \\
\hline}
		
\tablehead{\hline
\emph{Table \ref{bluefluxtab} cont'd.} & & & & & & \\ \hline
H II region position &  [O II]  &  H $\gamma$ &  H$\beta$  &  [O III]  &  [O III]  &  $R/R_{e}$  \\ 
& $\lambda3726 + \lambda3729$ & & & $\lambda4959$ & $\lambda5007$ & \\
\hline}
		
\tabletail{\hline}
\footnotesize
\begin{supertabular}{| l l l l l l l |}
\multicolumn{7}{| c |}{IC 5309} \\
-3.6, +0.0 & 97 $\pm$ 4 & $\le$ 3 & 100 $\pm$ 5 & 7 $\pm$ 4 & 16 $\pm$ 4 & 0.109 \\       
-1.9, +2.2 & 12 $\pm$ 8 & $\le$ 5 & 100 $\pm$ 5 & $\le$ 5 & 11 $\pm$ 5 & 0.118 \\      
+1.7, -2.0 & 58 $\pm$ 6 & $\le$ 5 & 100 $\pm$ 5 & $\le$ 3 & 16 $\pm$ 6 & 0.199 \\       
-0.1, -4.0 & 104 $\pm$ 9 & $\le$ 7 & 100 $\pm$ 6 & $\le$ 6 & 17 $\pm$ 5 & 0.222 \\      
-0.1, +4.2 & 67 $\pm$ 6 & $\le$ 6 & 100 $\pm$ 6 & 15 $\pm$ 4 & 14 $\pm$ 9 & 0.233 \\       
-5.6, -1.9 & 145 $\pm$ 7 & 47 $\pm$ 8 & 100 $\pm$ 5 & 11 $\pm$ 4 & 23 $\pm$ 4 & 0.240 \\       
+3.7, -0.1 & 118 $\pm$ 8 & $\le$ 6 & 100 $\pm$ 5 & $\le$ 6 & 7 $\pm$ 5 & 0.270 \\       
-5.3, +6.2 & 72 $\pm$ 5 & 49 $\pm$ 5 & 100 $\pm$ 4 & 11 $\pm$ 4 & 15 $\pm$ 4 & 0.384 \\       
-3.4, +8.3 & 76 $\pm$ 7 & 57 $\pm$ 5 & 100 $\pm$ 6 & $\le$ 4 & 11 $\pm$ 3 & 0.449 \\     
+7.1, -4.1 & 65 $\pm$ 6 & 47 $\pm$ 4 & 100 $\pm$ 4 & $\le$ 5 & 8 $\pm$ 4 & 0.500 \\      
-1.7,+10.4 & 133 $\pm$ 10 & 40 $\pm$ 6 & 100 $\pm$ 6 & 8 $\pm$ 6 & 12 $\pm$ 5 & 0.547 \\      
-7.0,+12.4 & 41 $\pm$ 10 & $\le$ 13 & 100 $\pm$ 6 & $\le$ 6 & 8 $\pm$ 6 & 0.714 \\      
+5.0,-14.4 & 194 $\pm$ 7 & 38 $\pm$ 6 & 100 $\pm$ 4 & 8 $\pm$ 3 & 12 $\pm$ 5 & 0.831 \\      
-8.6,+18.6 & 103 $\pm$ 12 & $\le$ 11 & 100 $\pm$ 9 & $\le$ 9 & 22 $\pm$ 9 & 1.048 \\      
+12.3,-14.5 & 175 $\pm$ 14 & 47 $\pm$ 7 & 100 $\pm$ 8 & 16 $\pm$ 5 & 30 $\pm$ 6 & 1.054 \\   
-16.0,+18.7 & 147 $\pm$ 13 & 23 $\pm$ 7 & 100 $\pm$ 8 & $\le$ 5 & 42 $\pm$ 5 & 1.244 \\    
-12.3,+22.9 & 151 $\pm$ 19 & $\le$ 11 & 100 $\pm$ 10 & 37 $\pm$ 10 & 41 $\pm$ 7 & 1.333 \\    
\hline
\multicolumn{7}{| c |}{NGC 7518} \\
-3.6, +0.1 & 78 $\pm$ 3 & 44 $\pm$ 5 & 100 $\pm$ 4 & 3 $\pm$ 4 & 14 $\pm$ 4 & 0.098 \\     
-1.6, +2.2 & 48 $\pm$ 4 & 36 $\pm$ 3 & 100 $\pm$ 4 & 4 $\pm$ 3 & 14 $\pm$ 3 & 0.100 \\     
+1.3, -2.0 & 125 $\pm$ 4 & 52 $\pm$ 6 & 100 $\pm$ 6 & 7 $\pm$ 3 & 25 $\pm$ 3 & 0.158 \\     
-10.8, +0.1 & 73 $\pm$ 8 & 42 $\pm$ 4 & 100 $\pm$ 7 & $\le$ 6 & 11 $\pm$ 6 & 0.423 \\   
-14.3,+10.6 & 97 $\pm$ 5 & 45 $\pm$ 3 & 100 $\pm$ 4 & $\le$ 4 & 25 $\pm$ 3 & 0.754 \\  
-18.6, +6.4 & 69 $\pm$ 13 & $\le$ 12 & 100 $\pm$ 10 & $\le$ 8 & 16 $\pm$ 7 & 0.830 \\   
+16.2, -8.4 & 95 $\pm$ 5 & 48 $\pm$ 4 & 100 $\pm$ 3 & $\le$ 3 & 15 $\pm$ 3 & 0.890 \\   
-19.2,+12.6 & 103 $\pm$ 3 & 53 $\pm$ 2 & 100 $\pm$ 2 & $\le$ 2 & 11 $\pm$ 2 & 0.987 \\  
+16.8,-14.6 & 142 $\pm$ 8 & $\le$ 7 & 100 $\pm$ 6 & $\le$ 7 & 15 $\pm$ 7 & 1.063 \\  
-23.7, +8.7 & 103 $\pm$ 7 & 43 $\pm$ 4 & 100 $\pm$ 4 & 4 $\pm$ 3 & 17 $\pm$ 4 & 1.085 \\   
+21.4,-16.6 & 131 $\pm$ 6 & 36 $\pm$ 5 & 100 $\pm$ 4 & $\le$ 4 & 17 $\pm$ 4 & 1.285 \\  
+33.6,-12.8 & 173 $\pm$ 13 & 29 $\pm$ 7 & 100 $\pm$ 8 & 5 $\pm$ 7 & 28 $\pm$ 5 & 1.698 \\  
+48.6,-19.1 & 138 $\pm$ 34 & $\le$ 25 & 100 $\pm$ 22 & $\le$ 22 & 13 $\pm$ 23 & 2.435 \\  
\hline
\multicolumn{7}{| c |}{NGC 7643} \\
-1.7,-10.1 & 109 $\pm$ 6 & 45 $\pm$ 4 & 100 $\pm$ 4 & 4 $\pm$ 3 & 11 $\pm$ 3 & 0.512 \\     
-7.0, +8.4 & 85 $\pm$ 7 & $\le$ 6 & 100 $\pm$ 6 & 6 $\pm$ 5 & 21 $\pm$ 5 & 0.547 \\      
+5.6,-10.2 & 79 $\pm$ 5 & 38 $\pm$ 3 & 100 $\pm$ 4 & $\le$ 3 & 5 $\pm$ 2 & 0.583 \\     
-5.9,+10.4 & 128 $\pm$ 10 & $\le$ 15 & 100 $\pm$ 11 & $\le$ 10 & 6 $\pm$ 6 & 0.598 \\     
-4.8,+12.4 & 86 $\pm$ 9 & $\le$ 7 & 100 $\pm$ 7 & 2 $\pm$ 6 & 5 $\pm$ 5 & 0.664 \\     
+4.8,-12.4 & 118 $\pm$ 6 & $\le$ 6 & 100 $\pm$ 5 & $\le$ 5 & 8 $\pm$ 4 & 0.664 \\     
+3.7,-14.3 & 105 $\pm$ 9 & $\le$ 7 & 100 $\pm$ 7 & $\le$ 6 & 10 $\pm$ 5 & 0.738 \\     
\hline
\multicolumn{7}{| c |}{NGC 7529} \\
+1.3, +6.1 & 180 $\pm$ 13 & $\le$ 8 & 100 $\pm$ 9 & 10 $\pm$ 6 & 29 $\pm$ 7 & 0.567 \\     
-1.6, -6.1 & 323 $\pm$ 6 & 43 $\pm$ 3 & 100 $\pm$ 4 & 21 $\pm$ 3 & 67 $\pm$ 3 & 0.574 \\       
-5.9, +6.2 & 138 $\pm$ 14 & $\le$ 8 & 100 $\pm$ 7 & 4 $\pm$ 4 & 17 $\pm$ 6 & 0.777 \\      
+10.7, -0.2 & 211 $\pm$ 6 & $\le$ 4 & 100 $\pm$ 3 & 11 $\pm$ 3 & 45 $\pm$ 3 & 0.974 \\      
-11.0, +0.2 & 232 $\pm$ 6 & 47 $\pm$ 3 & 100 $\pm$ 4 & 17 $\pm$ 3 & 61 $\pm$ 3 & 1.001 \\     
-13.1, +6.3 & 200 $\pm$ 3 & 49 $\pm$ 2 & 100 $\pm$ 1 & 16 $\pm$ 1 & 51 $\pm$ 1 & 1.322 \\     
+7.9, -12.4 & 242 $\pm$ 6 & 55 $\pm$ 4 & 100 $\pm$ 5 & 20 $\pm$ 4 & 67 $\pm$ 3 & 1.338 \\     
-8.2, +12.4 & 263 $\pm$ 9 & 82 $\pm$ 5 & 100 $\pm$ 4 & 42 $\pm$ 5 & 79 $\pm$ 6 & 1.353 \\     
+17.9, -0.3 & 353 $\pm$ 13 & 38 $\pm$ 6 & 100 $\pm$ 6 & 23 $\pm$ 5 & 97 $\pm$ 7 & 1.629 \\   
+15.1, -12.5 & 428 $\pm$ 8 & 56 $\pm$ 4 & 100 $\pm$ 5 & 14 $\pm$ 5 & 56 $\pm$ 5 & 1.785 \\     
+20.0, -6.4 & 449 $\pm$ 10 & 42 $\pm$ 5 & 100 $\pm$ 5 & 22 $\pm$ 5 & 58 $\pm$ 5 & 1.910 \\   
+29.6, -12.7 & 359 $\pm$ 16 & $\le$ 9 & 100 $\pm$ 7 & 38 $\pm$ 7 & 127 $\pm$ 7 &  2.924 \\     
\hline
\multicolumn{7}{| c |}{NGC 7591} \\
-0.8, -4.0 & 88 $\pm$ 6 & $\le$ 5 & 100 $\pm$ 6 & $\le$ 4 & 19 $\pm$ 4 & 0.328 \\     
-3.6, +0.0 & 113 $\pm$ 11 & $\le$ 10 & 100 $\pm$ 6 & 4 $\pm$ 3 & 31 $\pm$ 5 & 0.362 \\     
+2.9, +6.1 & 135 $\pm$ 7 & 51 $\pm$ 3 & 100 $\pm$ 4 & $\le$ 3 & 10 $\pm$ 3 & 0.447 \\    
-2.1, +8.3 & 129 $\pm$ 10 & $\le$ 8 & 100 $\pm$ 5 & 6 $\pm$ 4 & 11 $\pm$ 6 & 0.645 \\    
-8.8, +2.3 & 102 $\pm$ 4 & 36 $\pm$ 4 & 100 $\pm$ 4 & 6 $\pm$ 2 & 25 $\pm$ 4 & 0.750 \\   
+7.0,-10.2 & 91 $\pm$ 4 & 51 $\pm$ 5 & 100 $\pm$ 4 & 10 $\pm$ 3 & 18 $\pm$ 3 & 0.828 \\  
-5.0,+12.4 & 94 $\pm$ 5 & 30 $\pm$ 3 & 100 $\pm$ 2 & $\le$ 2 & 14 $\pm$ 2 & 0.998 \\    
+20.0, +1.9 & 123 $\pm$ 13 & $\le$ 7 & 100 $\pm$ 6 & 7 $\pm$ 5 & 16 $\pm$ 6 & 1.328 \\ 
-10.1,+14.6 & 100 $\pm$ 4 & 35 $\pm$ 3 & 100 $\pm$ 3 & 9 $\pm$ 2 & 15 $\pm$ 3 & 1.330 \\ 
-12.7,+18.6 & 143 $\pm$ 9 & 30 $\pm$ 5 & 100 $\pm$ 5 & 8 $\pm$ 4 & 27 $\pm$ 6 & 1.672 \\  
+35.0,-25.1 & 171 $\pm$ 9 & $\le$ 5 & 100 $\pm$ 4 & $\le$ 4 & 20 $\pm$ 4 & 2.990 \\  
+30.1, -2.4 & 272 $\pm$ 12 & 38 $\pm$ 8 & 100 $\pm$ 7 & 29 $\pm$ 9 & 97 $\pm$ 9 & 2.054 \\   
-28.5,+31.1 & 228 $\pm$ 11 & 52 $\pm$ 6 & 100 $\pm$ 6 & 13 $\pm$ 7 & 32 $\pm$ 7 & 3.085 \\  
+33.3,-33.2 & 186 $\pm$ 28 & $\le$ 13 & 100 $\pm$ 15 & 13 $\pm$ 11 & 63 $\pm$ 11 & 3.287 \\  
+42.2,-25.2 & 214 $\pm$ 6 & 33 $\pm$ 3 & 100 $\pm$ 3 & 2 $\pm$ 3 & 33 $\pm$ 3 & 3.421 \\  
+41.5,-19.0 & 323 $\pm$ 14 & 38 $\pm$ 7 & 100 $\pm$ 8 & $\le$ 8 & 66 $\pm$ 8 & 3.636 \\  
\hline

\end{supertabular}

\end{center}

\clearpage

\begin{center}

\tablecaption{Corrected emission line fluxes for H II region spectra using the VIRUS-P red setup.}
\label{redfluxtab}	

\tablefirsthead{\hline
H II region &  H $\beta$  &  [O III]  &  [O III] & [N II]  &  H $\alpha$  &  [N II] & [S II] & [S II]  &  $R/R_{e}$ & c  \\ 
postion (arcsec & & $\lambda4959$ & $\lambda5007$ & $\lambda6548$ & & $\lambda6583$ & $\lambda6716$ & $\lambda6731$ & & \\
from center) & & & & & & & & & & \\
\hline}
		
\tablehead{\hline
\emph{Table \ref{redfluxtab} cont'd.} & & & & & & & & & & \\ \hline
H II region &  H $\beta$  &  [O III]  &  [O III] & [N II]  &  H $\alpha$  &  [N II] & [S II] & [S II]  &  $R/R_{e}$ & c  \\ 
position& & $\lambda4959$ & $\lambda5007$ & $\lambda6548$ & & $\lambda6583$ & $\lambda6716$ & $\lambda6731$ & & \\
\hline}
		
\tabletail{\hline}
\footnotesize
\begin{supertabular}{| l l l l l l l l l l l |}
\multicolumn{11}{| c |}{IC 5309} \\
-3.6, +0.0 & 100 $\pm$ 6 & 8 $\pm$ 5 & 29 $\pm$ 5 & 30 $\pm$ 6 & 286 $\pm$ 5 & 90 $\pm$ 5 & 27 $\pm$ 8 & 28 $\pm$ 8 & 0.109 & 0.94 \\      
-1.9, +2.2 & 100 $\pm$ 6 & $\le$ 6 & 7 $\pm$ 6 & 259 $\pm$ 4 & 286 $\pm$ 4 & 965 $\pm$ 4 & 578 $\pm$ 4 & 452 $\pm$ 4 & 0.118 & 0.73 \\    
+1.7, -2.0 & 100 $\pm$ 7 & $\le$ 6 & 12 $\pm$ 7 & 20 $\pm$ 8 & 286 $\pm$ 8 & 90 $\pm$ 7 & 29 $\pm$ 8 & 24 $\pm$ 6 & 0.199 & 1.00 \\      
-0.1, -4.0 & 100 $\pm$ 5 & 4 $\pm$ 4 & 11 $\pm$ 5 & 22 $\pm$ 4 & 286 $\pm$ 4 & 86 $\pm$ 4 & 29 $\pm$ 4 & 26 $\pm$ 5 & 0.222 & 0.88 \\     
-0.1, +4.2 & 100 $\pm$ 6 & 3 $\pm$ 4 & 10 $\pm$ 4 & 27 $\pm$ 10 & 286 $\pm$ 10 & 88 $\pm$ 9 & 31 $\pm$ 12 & 27 $\pm$ 12 & 0.233 & 1.32 \\      
-5.6, -1.9 & 100 $\pm$ 4 & 2 $\pm$ 5 & 12 $\pm$ 5 & 18 $\pm$ 5 & 286 $\pm$ 5 & 76 $\pm$ 4 & 41 $\pm$ 6 & 28 $\pm$ 5 & 0.240 & 0.80 \\      
+3.7, -0.1 & 100 $\pm$ 7 & 4 $\pm$ 5 & 19 $\pm$ 6 & 25 $\pm$ 8 & 286 $\pm$ 8 & 95 $\pm$ 7 & 31 $\pm$ 7 & 35 $\pm$ 7 & 0.270 & 1.07 \\      
-5.3, +6.2 & 100 $\pm$ 5 & 4 $\pm$ 5 & 26 $\pm$ 6 & 24 $\pm$ 6 & 286 $\pm$ 6 & 94 $\pm$ 5 & 34 $\pm$ 6 & 23 $\pm$ 6 & 0.384 & 0.78 \\      
-3.4, +8.3 & 100 $\pm$ 6 & 7 $\pm$ 6 & 26 $\pm$ 5 & 27 $\pm$ 6 & 286 $\pm$ 6 & 99 $\pm$ 4 & 36 $\pm$ 7 & 25 $\pm$ 8 & 0.449 & 0.87 \\    
+7.1, -4.1 & 100 $\pm$ 4 & $\le$ 4 & 8 $\pm$ 5 & 22 $\pm$ 5 & 286 $\pm$ 5 & 106 $\pm$ 5 & 48 $\pm$ 4 & 30 $\pm$ 7 & 0.500 & 0.67 \\     
-1.7,+10.4 & 100 $\pm$ 6 & 12 $\pm$ 5 & 33 $\pm$ 6 & 21 $\pm$ 9 & 286 $\pm$ 9 & 70 $\pm$ 6 & 42 $\pm$ 8 & 45 $\pm$ 11 & 0.547 & 0.55 \\     
-7.0,+12.4 & 100 $\pm$ 8 & $\le$ 7 & 6 $\pm$ 9 & 22 $\pm$ 13 & 286 $\pm$ 13 & 91 $\pm$ 11 & 54 $\pm$ 11 & 48 $\pm$ 12 & 0.714 & 0.55 \\     
+5.0,-14.4 & 100 $\pm$ 4 & 1 $\pm$ 3 & 19 $\pm$ 4 & 19 $\pm$ 6 & 286 $\pm$ 6 & 77 $\pm$ 5 & 53 $\pm$ 4 & 41 $\pm$ 7 & 0.831 & 0.43 \\     
-8.6,+18.6 & 100 $\pm$ 8 & 41 $\pm$ 7 & 49 $\pm$ 7 & 23 $\pm$ 11 & 286 $\pm$ 11 & 90 $\pm$ 13 & 45 $\pm$ 17 & 29 $\pm$ 17 & 1.048 & 0.58 \\     
+12.3,-14.5 & 100 $\pm$ 6 & 13 $\pm$ 6 & 64 $\pm$ 7 & 16 $\pm$ 9 & 286 $\pm$ 9 & 82 $\pm$ 11 & 49 $\pm$ 10 & 54 $\pm$ 17 & 1.054 & 0.75 \\   
-16.0,+18.7 & 100 $\pm$ 10 & 10 $\pm$ 12 & 38 $\pm$ 10 & 51 $\pm$ 13 & 286 $\pm$ 13 & 84 $\pm$ 13 & 55 $\pm$ 13 & 108 $\pm$ 17 & 1.244 & 0.34 \\   
-12.3,+22.9 & 100 $\pm$ 14 & 10 $\pm$ 10 & 50 $\pm$ 10 & 19 $\pm$ 17 & 286 $\pm$ 17 & 49 $\pm$ 11 & 48 $\pm$ 13 & 67 $\pm$ 22 & 1.333 & 0.51 \\   
\hline
\multicolumn{11}{| c |}{NGC 7518} \\
-3.6, +0.1 & 100 $\pm$ 5 & 9 $\pm$ 3 & 21 $\pm$ 4 & 30 $\pm$ 5 & 286 $\pm$ 6 & 117 $\pm$ 3 & 33 $\pm$ 3 & 26 $\pm$ 4 & 0.098 & 0.90 \\     
-1.6, +2.2 & 100 $\pm$ 7 & $\le$ 7 & 25 $\pm$ 9 & 29 $\pm$ 7 & 286 $\pm$ 7 & 133 $\pm$ 7 & 64 $\pm$ 6 & 53 $\pm$ 6 & 0.100 & 0.29 \\    
+1.3, -2.0 & 100 $\pm$ 9 & $\le$ 8 & 30 $\pm$ 7 & 31 $\pm$ 11 & 286 $\pm$ 11 & 121 $\pm$ 11 & 31 $\pm$ 13 & 26 $\pm$ 17 & 0.158 & 1.75 \\    
-10.8, +0.1 & 100 $\pm$ 7 & $\le$ 6 & 9 $\pm$ 8 & 25 $\pm$ 6 & 286 $\pm$ 6 & 113 $\pm$ 7 & 30 $\pm$ 8 & 2 $\pm$ 12 & 0.423 & 0.53 \\  
-14.3,+10.6 & 100 $\pm$ 7 & $\le$ 6 & 8 $\pm$ 5 & 25 $\pm$ 6 & 286 $\pm$ 6 & 92 $\pm$ 11 & 18 $\pm$ 10 & 19 $\pm$ 7 & 0.754 & 1.38 \\ 
-18.6, +6.4 & 100 $\pm$ 9 & $\le$ 7 & 7 $\pm$ 8 & 23 $\pm$ 11 & 286 $\pm$ 11 & 98 $\pm$ 8 & 48 $\pm$ 9 & 30 $\pm$ 15 & 0.830 & 0.38 \\  
+16.2, -8.4 & 100 $\pm$ 8 & $\le$ 5 & 19 $\pm$ 7 & 26 $\pm$ 10 & 286 $\pm$ 10 & 90 $\pm$ 10 & 23 $\pm$ 9 & 25 $\pm$ 11 & 0.890 & 1.45 \\  
-19.2,+12.6 & 100 $\pm$ 4 & 4 $\pm$ 4 & 14 $\pm$ 3 & 37 $\pm$ 3 & 286 $\pm$ 3 & 104 $\pm$ 4 & 29 $\pm$ 3 & 18 $\pm$ 7 & 0.987 & 0.87 \\  
+16.8,-14.6 & 100 $\pm$ 12 & 36 $\pm$ 15 & 19 $\pm$ 15 & 13 $\pm$ 14 & 286 $\pm$ 14 & 128 $\pm$ 14 & 60 $\pm$ 17 & 38 $\pm$ 32 & 1.063 & 1.07 \\ 
-23.7, +8.7 & 100 $\pm$ 12 & $\le$ 14 & 5 $\pm$ 12 & 23 $\pm$ 11 & 286 $\pm$ 11 & 84 $\pm$ 14 & 19 $\pm$ 17 & 10 $\pm$ 11 & 1.085 & 1.15 \\  
+21.4,-16.6 & 100 $\pm$ 9 & $\le$ 7 & 12 $\pm$ 10 & 38 $\pm$ 9 & 286 $\pm$ 9 & 116 $\pm$ 9 & 52 $\pm$ 13 & 56 $\pm$ 16 & 1.285 & 0.61 \\ 
+33.6,-12.8 & 100 $\pm$ 8 & 18 $\pm$ 9 & 22 $\pm$ 9 & 29 $\pm$ 12 & 286 $\pm$ 12 & 104 $\pm$ 10 & 49 $\pm$ 8 & 37 $\pm$ 10 & 1.698 & 0.54 \\  
+48.6,-19.1 & 100 $\pm$ 22 & $\le$ 23 & 6 $\pm$ 23 & 17 $\pm$ 45 & 286 $\pm$ 45 & 114 $\pm$ 38 & $\le$ 47 & 67 $\pm$ 47 & 2.435 & 1.06 \\ 
\hline
\multicolumn{11}{| c |}{NGC 7643} \\
-1.7,-10.1 & 100 $\pm$ 7 & 8 $\pm$ 8 & 26 $\pm$ 6 & 28 $\pm$ 9 & 286 $\pm$ 9 & 103 $\pm$ 10 & 39 $\pm$ 8 & 31 $\pm$ 6 & 0.512 & 0.86 \\       
-7.0, +8.4 & 100 $\pm$ 10 & $\le$ 8 & 27 $\pm$ 13 & 45 $\pm$ 13 & 286 $\pm$ 13 & 124 $\pm$ 11 & 51 $\pm$ 15 & 25 $\pm$ 18 & 0.547 & 0.65 \\       
+5.6,-10.2 & 100 $\pm$ 6 & 5 $\pm$ 5 & 26 $\pm$ 6 & 32 $\pm$ 7 & 286 $\pm$ 7 & 115 $\pm$ 5 & 38 $\pm$ 8 & 35 $\pm$ 7 & 0.583 & 0.61 \\       
-5.9,+10.4 & 100 $\pm$ 12 & $\le$ 10 & 15 $\pm$ 9 & 27 $\pm$ 7 & 286 $\pm$ 7 & 127 $\pm$ 8 & 37 $\pm$ 7 & 48 $\pm$ 10 & 0.598 & 0.75 \\       
-4.8,+12.4 & 100 $\pm$ 9 & $\le$ 16 & 9 $\pm$ 11 & 69 $\pm$ 16 & 286 $\pm$ 16 & 165 $\pm$ 14 & 74 $\pm$ 11 & 37 $\pm$ 15 & 0.664 & -0.15 \\     
+4.8,-12.4 & 100 $\pm$ 6 & $\le$ 6 & 5 $\pm$ 4 & 32 $\pm$ 6 & 286 $\pm$ 6 & 113 $\pm$ 4 & 46 $\pm$ 3 & 43 $\pm$ 5 & 0.664 & 0.31 \\       
+3.7,-14.3 & 100 $\pm$ 6 & $\le$ 4 & $\le$ 6 & 32 $\pm$ 7 & 286 $\pm$ 7 & 107 $\pm$ 6 & 39 $\pm$ 8 & 49 $\pm$ 12 & 0.738 & 0.68 \\       
\hline
\multicolumn{11}{| c |}{IC 1474} \\
+3.3, +6.1 & 100 $\pm$ 12 & 13 $\pm$ 10 & 93 $\pm$ 10 & 32 $\pm$ 10 & 286 $\pm$ 10 & 97 $\pm$ 13 & 45 $\pm$ 10 & 55 $\pm$ 20 & 0.454 & 0.58 \\     
-1.6, +8.4 & 100 $\pm$ 5 & $\le$ 4 & 27 $\pm$ 3 & 35 $\pm$ 5 & 286 $\pm$ 5 & 106 $\pm$ 4 & 47 $\pm$ 4 & 35 $\pm$ 4 & 0.639 & 0.60 \\     
-6.4, +4.3 & 100 $\pm$ 11 & $\le$ 13 & 20 $\pm$ 10 & 31 $\pm$ 15 & 286 $\pm$ 15 & 107 $\pm$ 14 & 47 $\pm$ 12 & 30 $\pm$ 18 & 0.640 & 1.40 \\     
-8.7, +2.3 & 100 $\pm$ 3 & 4 $\pm$ 2 & 25 $\pm$ 2 & 28 $\pm$ 2 & 286 $\pm$ 2 & 99 $\pm$ 1 & 48 $\pm$ 1 & 34 $\pm$ 2 & 0.748 & 0.73 \\     
+8.4, -8.3 & 100 $\pm$ 9 & $\le$ 7 & 23 $\pm$ 7 & 16 $\pm$ 13 & 286 $\pm$ 13 & 107 $\pm$ 12 & 30 $\pm$ 12 & 28 $\pm$ 12 & 0.771 & 1.12 \\     
+6.0,-10.2 & 100 $\pm$ 3 & $\le$ 3 & 13 $\pm$ 3 & 22 $\pm$ 4 & 286 $\pm$ 4 & 89 $\pm$ 3 & 37 $\pm$ 3 & 29 $\pm$ 4 & 0.796 & 0.78 \\    
+13.0, -4.2 & 100 $\pm$ 2 & 8 $\pm$ 2 & 34 $\pm$ 2 & 27 $\pm$ 2 & 286 $\pm$ 2 & 90 $\pm$ 1 & 51 $\pm$ 2 & 38 $\pm$ 2 & 0.877 & 0.65 \\   
-8.8, +8.5 & 100 $\pm$ 3 & 5 $\pm$ 3 & 23 $\pm$ 3 & 33 $\pm$ 2 & 286 $\pm$ 2 & 100 $\pm$ 2 & 43 $\pm$ 2 & 33 $\pm$ 3 & 0.954 & 0.63 \\     
+15.6,-14.5 & 100 $\pm$ 3 & 9 $\pm$ 4 & 38 $\pm$ 4 & 25 $\pm$ 6 & 286 $\pm$ 6 & 97 $\pm$ 4 & 45 $\pm$ 4 & 24 $\pm$ 6 & 1.446 & 0.71 \\  
+18.0,-12.6 & 100 $\pm$ 5 & 8 $\pm$ 6 & 41 $\pm$ 5 & 25 $\pm$ 7 & 286 $\pm$ 7 & 92 $\pm$ 6 & 51 $\pm$ 6 & 36 $\pm$ 12 & 1.486 & 0.81 \\  
+30.3,-27.0 & 100 $\pm$ 22 & 23 $\pm$ 22 & 100 $\pm$ 22 & 39 $\pm$ 23 & 286 $\pm$ 23 & 58 $\pm$ 17 & 24 $\pm$ 24 & 17 $\pm$ 24 & 2.821 & 0.75 \\  
+35.0,-22.9 & 100 $\pm$ 10 & 29 $\pm$ 6 & 102 $\pm$ 6 & 30 $\pm$ 10 & 286 $\pm$ 10 & 87 $\pm$ 5 & 33 $\pm$ 8 & 25 $\pm$ 16 & 2.897 & 0.75 \\  
\hline
\multicolumn{11}{| c |}{NGC 7529} \\
-2.2, +2.2 & 100 $\pm$ 7 & 15 $\pm$ 7 & 31 $\pm$ 6 & 28 $\pm$ 6 & 286 $\pm$ 6 & 102 $\pm$ 4 & 50 $\pm$ 3 & 35 $\pm$ 14 & 0.285 & 0.34 \\     
+0.1, -4.0 & 100 $\pm$ 8 & 5 $\pm$ 6 & 29 $\pm$ 6 & 25 $\pm$ 5 & 286 $\pm$ 5 & 110 $\pm$ 6 & 76 $\pm$ 5 & 47 $\pm$ 9 & 0.364 & 0.42 \\      
+1.3, +6.1 & 100 $\pm$ 13 & 16 $\pm$ 10 & 47 $\pm$ 10 & 23 $\pm$ 7 & 286 $\pm$ 7 & 89 $\pm$ 9 & 82 $\pm$ 14 & 75 $\pm$ 14 & 0.567 & 0.39 \\      
-1.6, -6.1 & 100 $\pm$ 3 & 11 $\pm$ 4 & 40 $\pm$ 4 & 22 $\pm$ 4 & 286 $\pm$ 4 & 87 $\pm$ 4 & 54 $\pm$ 4 & 51 $\pm$ 5 & 0.574 & 0.53 \\       
-7.1, -3.9 & 100 $\pm$ 3 & 21 $\pm$ 3 & 62 $\pm$ 3 & 23 $\pm$ 3 & 286 $\pm$ 3 & 90 $\pm$ 4 & 51 $\pm$ 3 & 44 $\pm$ 2 & 0.737 & 0.34 \\      
+7.3, -4.1 & 100 $\pm$ 5 & $\le$ 3 & 28 $\pm$ 4 & 31 $\pm$ 3 & 286 $\pm$ 3 & 90 $\pm$ 4 & 68 $\pm$ 5 & 49 $\pm$ 9 & 0.762 & 0.41 \\      
-5.9, +6.2 & 100 $\pm$ 10 & 22 $\pm$ 11 & 31 $\pm$ 9 & 23 $\pm$ 8 & 286 $\pm$ 8 & 118 $\pm$ 12 & 55 $\pm$ 10 & 44 $\pm$ 10 & 0.777 & 0.70 \\       
-3.3, -8.1 & 100 $\pm$ 3 & 24 $\pm$ 2 & 83 $\pm$ 2 & 18 $\pm$ 3 & 286 $\pm$ 3 & 70 $\pm$ 3 & 57 $\pm$ 2 & 42 $\pm$ 3 & 0.794 & 0.31 \\       
+9.0, -2.1 & 100 $\pm$ 4 & 17 $\pm$ 4 & 33 $\pm$ 4 & 29 $\pm$ 5 & 286 $\pm$ 5 & 88 $\pm$ 4 & 58 $\pm$ 5 & 35 $\pm$ 5 & 0.837 & 0.35 \\      
-2.7,+10.3 & 100 $\pm$ 6 & 26 $\pm$ 6 & 73 $\pm$ 6 & 10 $\pm$ 3 & 286 $\pm$ 5 & 76 $\pm$ 5 & 60 $\pm$ 7 & 39 $\pm$ 16 & 0.968 & 0.07 \\     
+10.7, -0.2 & 100 $\pm$ 3 & 23 $\pm$ 3 & 66 $\pm$ 4 & 21 $\pm$ 3 & 286 $\pm$ 3 & 71 $\pm$ 3 & 67 $\pm$ 6 & 54 $\pm$ 5 & 0.974 & 0.43 \\     
-11.0, +0.2 & 100 $\pm$ 3 & 21 $\pm$ 2 & 57 $\pm$ 3 & 19 $\pm$ 1 & 286 $\pm$ 1 & 80 $\pm$ 2 & 47 $\pm$ 3 & 42 $\pm$ 3 & 1.001 & 0.61 \\     
-12.7, -1.8 & 100 $\pm$ 3 & 11 $\pm$ 3 & 61 $\pm$ 4 & 25 $\pm$ 4 & 286 $\pm$ 4 & 67 $\pm$ 4 & 56 $\pm$ 4 & 44 $\pm$ 12 & 1.162 & 0.16 \\     
+9.4,-10.2 & 100 $\pm$ 4 & 16 $\pm$ 4 & 47 $\pm$ 2 & 24 $\pm$ 3 & 286 $\pm$ 3 & 94 $\pm$ 4 & 60 $\pm$ 3 & 36 $\pm$ 2 & 1.261 & 0.35 \\     
-13.1, +6.3 & 100 $\pm$ 5 & 6 $\pm$ 4 & 49 $\pm$ 4 & 25 $\pm$ 3 & 286 $\pm$ 3 & 82 $\pm$ 4 & 56 $\pm$ 4 & 51 $\pm$ 5 & 1.322 & 0.68 \\     
+7.9, -12.4 & 100 $\pm$ 6 & 14 $\pm$ 6 & 73 $\pm$ 5 & 24 $\pm$ 7 & 286 $\pm$ 7 & 75 $\pm$ 5 & 71 $\pm$ 8 & 69 $\pm$ 10 & 1.338 & 0.43 \\     
-8.2, +12.4 & 100 $\pm$ 6 & 21 $\pm$ 8 & 86 $\pm$ 7 & 24 $\pm$ 9 & 286 $\pm$ 9 & 86 $\pm$ 9 & 64 $\pm$ 9 & 44 $\pm$ 11 & 1.353 & 0.53 \\     
+14.5, -4.2 & 100 $\pm$ 5 & 21 $\pm$ 3 & 58 $\pm$ 3 & 21 $\pm$ 3 & 286 $\pm$ 3 & 83 $\pm$ 3 & 80 $\pm$ 5 & 62 $\pm$ 5 & 1.373 & 0.39 \\    
+16.2, -2.2 & 100 $\pm$ 7 & 30 $\pm$ 6 & 98 $\pm$ 6 & 20 $\pm$ 8 & 286 $\pm$ 8 & 64 $\pm$ 8 & 64 $\pm$ 6 & 36 $\pm$ 12 & 1.483 & 0.36 \\    
+17.9, -0.3 & 100 $\pm$ 17 & 87 $\pm$ 15 & 228 $\pm$ 15 & 29 $\pm$ 7 & 286 $\pm$ 7 & 76 $\pm$ 13 & 56 $\pm$ 20 & 28 $\pm$ 17 & 1.629 & 0.73 \\     
+16.6,-10.3 & 100 $\pm$ 3 & 23 $\pm$ 3 & 94 $\pm$ 2 & 25 $\pm$ 1 & 286 $\pm$ 1 & 82 $\pm$ 2 & 59 $\pm$ 3 & 38 $\pm$ 7 & 1.776 & 0.51 \\   
+15.1, -12.5 & 100 $\pm$ 3 & 20 $\pm$ 1 & 59 $\pm$ 2 & 26 $\pm$ 2 & 286 $\pm$ 2 & 77 $\pm$ 2 & 49 $\pm$ 1 & 33 $\pm$ 3 & 1.785 & 0.36 \\     
-13.9,+14.6 & 100 $\pm$ 7 & 38 $\pm$ 7 & 100 $\pm$ 7 & 32 $\pm$ 4 & 286 $\pm$ 4 & 65 $\pm$ 6 & 77 $\pm$ 8 & 45 $\pm$ 17 & 1.832 & 0.22 \\  
+18.4, -8.4 & 100 $\pm$ 4 & 21 $\pm$ 4 & 97 $\pm$ 4 & 18 $\pm$ 5 & 286 $\pm$ 5 & 82 $\pm$ 5 & 64 $\pm$ 6 & 30 $\pm$ 11 & 1.835 & 0.27 \\    
+1.2,-20.4 & 100 $\pm$ 12 & 72 $\pm$ 9 & 206 $\pm$ 9 & $\le$ 12 & 286 $\pm$ 12 & 26 $\pm$ 8 & 49 $\pm$ 8 & 61 $\pm$ 25 & 1.858 & -0.09 \\   
+20.0, -6.4 & 100 $\pm$ 19 & $\le$ 20 & 119 $\pm$ 16 & $\le$ 21 & 286 $\pm$ 21 & 65 $\pm$ 21 & 71 $\pm$ 21 & 48 $\pm$ 59 & 1.910 & 1.33 \\     
-21.5, -3.7 & 100 $\pm$ 5 & 72 $\pm$ 9 & 189 $\pm$ 9 & $\le$ 7 & 286 $\pm$ 7 & 24 $\pm$ 7 & 52 $\pm$ 10 & 20 $\pm$ 14 & 1.985 & 0.17 \\    
+18.8,-16.5 & 100 $\pm$ 7 & 9 $\pm$ 7 & 47 $\pm$ 8 & 25 $\pm$ 10 & 286 $\pm$ 7 & 73 $\pm$ 7 & 82 $\pm$ 8 & 36 $\pm$ 10 & 2.274 & 0.43 \\   
+23.8,-10.4 & 100 $\pm$ 7 & 29 $\pm$ 10 & 138 $\pm$ 10 & 17 $\pm$ 5 & 286 $\pm$ 5 & 76 $\pm$ 9 & 51 $\pm$ 9 & 37 $\pm$ 19 & 2.362 & 0.47 \\   
+29.6, -12.7 & 100 $\pm$ 7 & 16 $\pm$ 7 & 81 $\pm$ 7 & 24 $\pm$ 7 & 286 $\pm$ 7 & 46 $\pm$ 7 & 93 $\pm$ 12 & 33 $\pm$ 15 & 2.924 & 0.31 \\     
\hline
\multicolumn{11}{| c |}{NGC 7591} \\
-0.8, -4.0 & 100 $\pm$ 5 & 4 $\pm$ 4 & 13 $\pm$ 4 & 30 $\pm$ 2 & 286 $\pm$ 2 & 112 $\pm$ 3 & 27 $\pm$ 4 & 16 $\pm$ 4 & 0.328 & 0.78 \\      
-3.6, +0.0 & 100 $\pm$ 8 & 7 $\pm$ 8 & 27 $\pm$ 7 & 41 $\pm$ 9 & 286 $\pm$ 9 & 252 $\pm$ 5 & 40 $\pm$ 4 & 30 $\pm$ 8 & 0.362 & 0.58 \\     
+2.9, +6.1 & 100 $\pm$ 8 & 6 $\pm$ 5 & 22 $\pm$ 5 & 24 $\pm$ 3 & 286 $\pm$ 3 & 106 $\pm$ 5 & 43 $\pm$ 5 & 33 $\pm$ 20 & 0.447 & 1.21 \\    
-2.1, +8.3 & 100 $\pm$ 4 & $\le$ 3 & 17 $\pm$ 3 & 25 $\pm$ 3 & 286 $\pm$ 3 & 96 $\pm$ 3 & 36 $\pm$ 3 & 25 $\pm$ 6 & 0.645 & 1.32 \\     
-8.8, +2.3 & 100 $\pm$ 6 & $\le$ 5 & 23 $\pm$ 5 & 29 $\pm$ 6 & 286 $\pm$ 6 & 125 $\pm$ 5 & 42 $\pm$ 4 & 29 $\pm$ 11 & 0.750 & 0.83 \\     
+7.0,-10.2 & 100 $\pm$ 5 & $\le$ 5 & 22 $\pm$ 5 & 26 $\pm$ 2 & 286 $\pm$ 2 & 98 $\pm$ 5 & 32 $\pm$ 5 & 22 $\pm$ 17 & 0.828 & 1.32 \\    
-5.0,+12.4 & 100 $\pm$ 2 & $\le$ 2 & 21 $\pm$ 2 & 29 $\pm$ 2 & 286 $\pm$ 2 & 99 $\pm$ 2 & 35 $\pm$ 2 & 24 $\pm$ 7 & 0.998 & 1.30 \\    
+20.0, +1.9 & 100 $\pm$ 4 & 9 $\pm$ 5 & 36 $\pm$ 5 & 28 $\pm$ 6 & 286 $\pm$ 6 & 89 $\pm$ 6 & 43 $\pm$ 6 & 22 $\pm$ 6 & 1.328 & 1.03 \\   
-10.1,+14.6 & 100 $\pm$ 3 & $\le$ 2 & 15 $\pm$ 3 & 26 $\pm$ 1 & 286 $\pm$ 1 & 90 $\pm$ 2 & 38 $\pm$ 2 & 27 $\pm$ 10 & 1.330 & 1.12 \\  
-12.7,+18.6 & 100 $\pm$ 5 & 12 $\pm$ 5 & 27 $\pm$ 5 & 28 $\pm$ 3 & 286 $\pm$ 3 & 104 $\pm$ 5 & 55 $\pm$ 7 & 29 $\pm$ 17 & 1.672 & 1.18 \\  
+30.1, -2.4 & 100 $\pm$ 19 & $\le$ 19 & 44 $\pm$ 19 & $\le$ 25 & 286 $\pm$ 25 & 116 $\pm$ 25 & 77 $\pm$ 19 & 22 $\pm$ 19 & 2.054 & 0.15 \\   
+35.0,-25.1 & 100 $\pm$ 5 & $\le$ 3 & 17 $\pm$ 5 & 31 $\pm$ 5 & 286 $\pm$ 5 & 96 $\pm$ 6 & 47 $\pm$ 6 & 29 $\pm$ 16 & 2.990 & 1.16 \\  
-28.5,+31.1 & 100 $\pm$ 8 & $\le$ 8 & 50 $\pm$ 8 & 18 $\pm$ 6 & 286 $\pm$ 6 & 95 $\pm$ 6 & 48 $\pm$ 8 & 27 $\pm$ 17 & 3.085 & 0.43 \\  
+33.3,-33.2 & 100 $\pm$ 17 & $\le$ 17 & 53 $\pm$ 13 & 57 $\pm$ 22 & 286 $\pm$ 22 & 94 $\pm$ 23 & 133 $\pm$ 17 & 12 $\pm$ 29 & 3.287 & 0.25 \\  
+42.2,-25.2 & 100 $\pm$ 4 & 16 $\pm$ 3 & 44 $\pm$ 3 & 24 $\pm$ 5 & 286 $\pm$ 5 & 91 $\pm$ 5 & 61 $\pm$ 5 & 35 $\pm$ 11 & 3.421 & 0.79 \\  
+41.5,-19.0 & 100 $\pm$ 9 & 12 $\pm$ 9 & 102 $\pm$ 9 & 27 $\pm$ 8 & 286 $\pm$ 8 & 75 $\pm$ 12 & 70 $\pm$ 12 & 42 $\pm$ 24 & 3.636 & 1.41 \\  
\hline
\end{supertabular}

\end{center}

\begin{table}
\begin{center}
\begin{tabular}{| c | c | c | c | c |}

\hline\hline & & & \\
(1) & (2) & (3) & (4) \\
Galaxy & DEF & Mean $f_{[O III]}/f_{H \beta}$ & Mean 12 + log(O/H) & (O/H) Gradient (dex/$R_e$) \\ \hline
IC 5309 & 0.29 & 0.59 $\pm$ 0.15 & 9.15 $\pm$ 0.06 & -0.13 $\pm$ 0.04 \\ \hline
NGC 7518 & 0.27 & 0.21 $\pm$ 0.10 & 9.23 $\pm$ 0.03 & -0.05 $\pm$ 0.02 \\ \hline
NGC 7643 & 0.65 & 0.17 $\pm$ 0.13 & 9.25 $\pm$ 0.02 & $\cdots$ \\ \hline
IC 1474 & -0.05 & 0.44 $\pm$ 0.18 & $\cdots$ & $\cdots$ \\ \hline
NGC 7529 & -0.08 & 0.88 $\pm$ 0.33 & 9.00 $\pm$ 0.09 & -0.17 $\pm$ 0.05 \\ \hline
NGC 7591 & -0.25 & 0.38 $\pm$ 0.11 & 9.14 $\pm$ 0.05 & -0.07 $\pm$ 0.02 \\ \hline
\end{tabular}
\caption{Mean oxygen data for our targets.}
\label{logotab}
\end{center}
\end{table}

\begin{table}
\begin{center}
\begin{tabular}{| c | l | l | l | l | l |}

\hline
Deficiency Group\tablenotemark{4} & Galaxy & $V_C$ (km/s) & $M_B$ & DEF\tablenotemark{5} & 12 + log(O/H)\tablenotemark{6} \\ \hline
\multicolumn{6}{| l |}{\emph{Virgo Cluster}} \\ \hline
 	     & NGC 4303 & 216 & -21.13 & -0.06 & 9.09 $\pm$ 0.02 \\
H I Normal   & NGC 4651 & 250 & -20.18 & ... & 8.99 $\pm$ 0.06 \\
    	     & NGC 4713 & 137 & -19.10 & ... & 8.84 $\pm$ 0.03 \\ \hline
	     & NGC 4254 & 250 & -20.95 & 0.01 & 9.18 $\pm$ 0.02 \\
Intermediate & NGC 4321 & 201 & -21.29 & 0.53 & 9.23 $\pm$ 0.02 \\
	     & NGC 4654 & 198 & -20.71 & -0.29 & 9.01 $\pm$ 0.03 \\ \hline
	     & NGC 4501 & 278 & -21.57 & 0.55 & 9.32 $\pm$ 0.05 \\
H I Deficient& NGC 4571 & 165 & -19.43 & 0.54 & 9.24 $\pm$ 0.02 \\
    	     & NGC 4689 & 185 & -19.86 & 0.90 & 9.28 $\pm$ 0.02 \\ \hline
\multicolumn{6}{| l |}{\emph{Field Galaxies from \citet{zaritsky94}}} \\ \hline
	     & NGC 628 & 107 & -20.32 & -0.01 & 8.94 $\pm$ 0.19 \\
	     & NGC 2903 & 228 & -19.85 & 0.33 & 9.12 $\pm$ 0.08 \\
	     & NGC 3521 & 268 & -19.88 & 0.06 & 8.97 $\pm$ 0.14 \\
	     & NGC 4258 & 234 & -20.59 & 0.25 & 8.97 $\pm$ 0.06 \\
	     & NGC 4736 & 209 & -19.37 & 0.68 & 9.01 $\pm$ 0.17 \\
N/A      & NGC 5033 & 251 & -21.03 & 0.19 & 8.84 $\pm$ 0.16 \\
	     & NGC 5055 & 242 & -20.14 & 0.12 & 9.21 $\pm$ 0.25 \\
	     & NGC 5194 & 250 & -20.75 & 0.12 & 9.23 $\pm$ 0.12 \\
	     & NGC 5457 & 190 & -20.45 & 0.28 & 8.52 $\pm$ 0.06 \\
	     & NGC 6946 & 240 & -20.78 & -0.45 & 9.06 $\pm$ 0.17 \\
	     & NGC 7331 & 282 & -21.10 & -0.04 & 9.03 $\pm$ 0.19 \\ \hline

\end{tabular}
\caption{Inclination-corrected circular velocities, absolute B magnitudes, H I deficiencies, and abundances for spiral galaxies in the Virgo cluster and in the field.}
\label{virgofieldtab}
\end{center}
\end{table}


\clearpage

\begin{thebibliography}{}
\expandafter\ifx\csname natexlab\endcsname\relax\def\natexlab#1{#1}\fi
%
\bibitem[Binggeli et al.(1985)]{binggeli85} Binggeli, B. \etal\ 1985, \aj, 90, 1681
%
\bibitem[Boselli \& Gavazzi(2006)]{boselli06} Boselli, A. \& Gavazzi, G. \ 2006, \pasp, 118, 517
%
\bibitem[Canizares et al.(1986)]{canizares86} Canizares, C.~R. \etal\ 1986, \apj, 304, 312
%
\bibitem[Cooper et al.(2008)]{cooper08} Cooper, M.~C. \etal\ 2008, \mnras, 390, 245
%
\bibitem[Dors \& Copetti(2006)]{dors06} Dors, O.~L., \& Copetti, M.~V.~F., \ 2006, \aap, 452, 473
%
\bibitem[Ellison et al.(2008)]{ellison08} Ellison, S.~L. \etal\ 2008, \aj, 135, 1877
%
\bibitem[Ellison et al.(2009)]{ellison09} Ellison, S.~L. \etal\ 2009, \mnras, 396, 1257
%
\bibitem[Finlator \& Dav\'{e}(2008)]{finlator08} Finlator, K., \& Dav\'{e}, R., \ 2008, 385, 2181
%
\bibitem[Fumagalli et al.(2009)]{fumagalli09} Fumagalli, M. \etal\ 2009, \apj, 697, 1811
%
\bibitem[Garnett et al.(1997)]{garnett97} Garnett, D.~R. \etal\ 1997, \apj, 489, 63
%
\bibitem[Gavazzi et al.(2005)]{gavazzi05} Gavazzi, G. \etal\ 2005, \aap, 2005, 429, 439
%
\bibitem[Gunn \& Gott(1972)]{gunn72} Gunn, J.~E., \& Gott, J.~R. \ 1972, \apj, 176, 1
%
\bibitem[Haynes et al.(1984)]{haynes84} Haynes, M.~P. \etal\ 1984, \araa, 22, 445
%
\bibitem[Henry et al.(1994)]{henry94} Henry, R.~B.~C. \etal\ 1994, \mnras, 266, 421
%
\bibitem[Hill et al.(2008)]{hill08} Hill, G.~J. \etal\ 2008, SPIE, 7014, 231
%
\bibitem[Kennicutt et al.(1993)]{kennicutt93} Kennicut, R.~C. \etal\ 2008, MxAA, 27, 21
%
\bibitem[Kewley \& Dopita(2002)]{kewley02} Kewley, L.~J. \& Dopita, M.~A., \ 2002, \apjs, 142, 35
%
\bibitem[Kewley et al.(2006)]{kewley06} Kewley, L.~J. \etal\ 2006, \aj, 131, 2004
%
\bibitem[Kong et al.(2002)]{kong02} Kong, X. \etal\ 2002, \aap, 396, 503
%
\bibitem[Levy et al.(2007)]{levy07} Levy, L. \etal\ 2007, \aj, 133, 1104
%
\bibitem[McGaugh(1991)]{mcgaugh91} McGaugh, S.~S. \ 1991, \apj, 380, 140
%
\bibitem[Mouhcine et al.(2007)]{mouhcine07} Mouhcine, M. \etal\ 2007, \mnras, 382, 801
%
\bibitem[Moustakas \& Kennicutt(2006)]{moustakas06} Moustakas, J. \& Kennicut, R.~C., \ 2006, \apj, 651, 155
%
\bibitem[Osterbrock \& Ferland(2006)]{osterbrock06} Osterbrock, D.~E. \& Ferland, G.~J. \ 2006, \emph{Astrophysics of Gaseous Nebulae and Active Galactic Nuclei} (2nd ed.; Sausalito, CA: University Science Books)
%
\bibitem[Pagel \& Patchett(1975)]{pagel75} Pagel, B.~E.~J. \& Patchett, B.~E., \ 1975, \mnras, 172, 13
%
\bibitem[Paturel et al.(2003)]{paturel03} Paturel, G. \etal\ 2003, \aap, 412, 45
%
\bibitem[Petropoulo et al.(2011)]{petropoulo11} Petropoulo, V. \etal\ 2011, \apj, 734, 32
%
\bibitem[Rose et al.(2010)]{rose10} Rose, J.~A. \etal\ 2010, \aj, 139, 765
%
\bibitem[Shields et al.(1991)]{shields91} Shields, G.~A. \etal\ 1991, \apj, 371, 82
%
\bibitem[Skillman et al.(1996)]{skillman96} Skillman, E.~D. \etal\ 1996, \apj, 462, 147
%
\bibitem[Solanes et al.(2001)]{solanes01} Solanes, J.~M. \etal\ 2001, \apj, 548, 97
%
\bibitem[Tremonti et al.(2004)]{tremonti04} Tremonti, C.~A. \etal\ 2004, \apj, 613, 898
%
\bibitem[Vigroux et al.(1989)]{vigroux89} Vigroux, L. \etal\ 1989, \aj, 98, 2044
%
\bibitem[Zaritsky et al.(1994)]{zaritsky94} Zaritsky, D. \etal\ 1994, \apj, 420, 87
%
\bibitem[Zhang et al.(2009)]{zhang09} Zhang, W. \etal\ 2009, \mnras, 397, 1243
%
\end{thebibliography}
\end{document}